\documentstyle[dina4,12pt,fleqn,psfig,rotate]{article}
\newcounter{subeq}
\renewcommand{\baselinestretch}{1.0}

\newcommand{\subeqno}{\stepcounter{subeq} \addtocounter{equation}{-1}}
\newcommand{\alabel}[1]{\addtocounter{equation}{-1}%
                        \refstepcounter{equation}%
                        \label{#1}%
                       }
\newcommand{\subeqres}{\setcounter{subeq}{0}}
\newcommand{\eqnoinc}{\addtocounter{equation}{1}}
\newcommand{\be}{\begin{equation}}
\newcommand{\ee}{\end{equation}}
\newcommand{\bea}{\begin{eqnarray}}
\newcommand{\eea}{\end{eqnarray}}
\newcommand{\beano}{\begin{eqnarray*}}
\newcommand{\eeano}{\end{eqnarray*}}
\newcommand{\bfr}{\begin{flushright}}
\newcommand{\efr}{\end{flushright}}

\begin{document}
\title{Two-body correlations in pionic systems\footnote{supported
 by DFG, BMFT, GSI Darmstadt and KFA J\"ulich}}
\author{J. M. H\"auser, W. Cassing and A. Peter\\
Institut f\"ur Theoretische Physik, Universit\"at Gie\ss en \\
35392 Gie\ss en, Germany \\}
\maketitle
\begin{abstract}
In close analogy to fermionic many-body theory the truncation
of the bosonic BBGKY density matrix hierarchy on the two-body level leads
to a coupled set of nonlinear equations of motion for the one-body density
matrix and the two-body correlation function.
These equations provide a nonperturbative
description of the nonequilibrium time evolution of
particle number conserving bosonic many-body systems including
the dynamical resummation of parquet-like diagrams.
Within this framework we study the
properties of a pionic system as a function of temperature and density
with focus on two-body quantities. For each temperature we find a related
pion density for which the relative strength of the two-body correlation
function assumes a maximum and the pionic system is far from the mean-field
limit. Since these correlated phases up to $T$=200 MeV only appear at
rather low pion density, the hot and dense pion gas as generated in
ultrarelativistic nucleus-nucleus collisions should be well described within
mean-field theory; i.e. the HBT analysis of pion sources from $\pi-\pi$
correlations should remain valid even in the case of strongly interacting
pions.
\end{abstract}
\newpage
\section{Introduction}
\label{introduction}

The theoretical description of strongly interacting hadronic many-body
systems in general requires nonperturbative methods. A systematic way
for constructing a wide class of such nonperturbative approaches is given
by the truncation of Green function hierarchies as obtained by inserting the
cluster expansion in terms of the corresponding connected Green functions.
In the lowest order truncation scheme one then ends up with the mean-field
level, i.e. TDHF\footnote{{\bf T}ime {\bf D}ependent {\bf H}artree
{\bf F}ock}. The next order truncation scheme, i.e.
neglecting the connected six-point function (the three-body correlation
function) as well as all higher order correlation functions leads to the
equations of two-body correlation dynamics, which in the fermionic case
have been denoted by NQCD\footnote{{\bf N}uclear {\bf Q}uantum
{\bf C}orrelation {\bf D}ynamics}.
The NQCD-method has already been successfully applied to the most important
fermionic many-body problem in hadronic physics, the nuclear many-body
problem \cite{1}-\cite{9}.
It has been shown to guarantee a simultaneous resummation of ring- and
ladder-diagrams in vertical and horizontal direction (a parquet resummation)
and thus to adequately take into account long- as well as short-range
nucleon-nucleon correlations \cite{5}.

It is the purpose of the present paper, to develop a related nonperturbative
theory for bosonic systems.
The most important bosonic many-body problem on the hadron level is that
of a pion gas, which can be generated e.g. in an ultrarelativistic
heavy-ion collision \cite{10}-\cite{12}.
Here especially the Hanbury Brown-Twiss (HBT) analysis of pion sources from
$\pi-\pi$ correlations \cite{13}-\cite{22}
might become questionable due to the strong
interaction between the pions.
In this work we will apply our method to study the properties
of two-body quantities in such systems.

The paper is organized as follows: In Sect. \ref{eqom} we will derive
the equations of motion of two-body boson correlation dynamics
starting from the Heisenberg equations
of motion for an arbitrary, particle number
conserving many-body Hamiltonian with a two-body interaction.
In Sect. \ref{diagramanalysis} we will investigate the topological structure
of the equations and analyze the various interaction terms diagrammatically.

Sect. \ref{partial} to Sect. \ref{relative} are devoted to a presentation
of numerical results for pionic systems, i.e. the comparison of different
limiting cases for the resummation of diagrams (Sect. \ref{partial}),
the computation of correlated near-equilibrium states
(Sect. \ref{generation}), the effect of correlations on two-body quantities
(Sect. \ref{coordinate}) and signatures for a phase transition of the system
as a function of density and temperature (Sect. \ref{relative}).
We will shift the specification
of the Hamiltonian density for interacting pions as well as the final
explicit equations of motion to the Appendices \ref{specification} and
\ref{eqomiso}, respectively.
\section{The equations of motion}
\label{eqom}
In this Sect. we develop the mathematical apparatus of correlation
dynamics for bosonic many-particle systems described by an arbitrary,
particle number conserving Hamiltonian with a two-body interaction.

In this respect, we first introduce the formalism of density matrices:
let $a_{\alpha}$ and $a_{\alpha}^\dagger$ be the creation and annihilation
operators for a system of bosons, where $\alpha$ stands for
a complete set of quantum numbers characterizing an element of an arbitrary,
orthonormal basis of the corresponding one-body Hilbert space,
with canonical equal-time commutation relations
\bea
[ a_{\alpha},a_{\beta}^\dagger ]
= \delta_{\alpha\beta} \; , \; \;
[ a_{\alpha},a_{\beta} ] = [ a_{\alpha}^\dagger,
a_{\beta}^\dagger ] = 0 \; .
\label{kommurel}
\eea
The matrix elements of the n-body density matrix are then defined by
\bea
{(\rho_n)}_{\alpha_1...\alpha_n\alpha_1'...\alpha_n'} =
{\langle}a_{\alpha_1'}^\dagger...a_{\alpha_n'}^\dagger a_{\alpha_n}...
a_{\alpha_1}{\rangle}
\; ,
\label{dmdefinition}
\eea
i.e., by the expectation value\footnote{in general an ensemble average with
an arbitrary nonequilibrium quantum statistical density operator}
of the normal-ordered operator product
of the corresponding creation and annihilation operators, all of these
considered at equal-time $t$ in the Heisenberg picture.
The matrix elements of the density matrices can thus be considered as
equal-time Green functions.

{}From the canonical commutation relations (\ref{kommurel}) we immediately
obtain the symmetry relations
\beano
{(\rho_1)}_{\alpha\alpha'}={(\rho_1)}_{\alpha'\alpha}^* \; , \; \;
{(\rho_2)}_{\alpha\beta\alpha'\beta'}={(\rho_2)}_{\alpha'\beta'\alpha\beta}^*
\; ,
\eeano
\bea
{(\rho_2)}_{\alpha\beta\alpha'\beta'}={(\rho_2)}_{\beta\alpha\alpha'\beta'}
={(\rho_2)}_{\alpha\beta\beta'\alpha'}={(\rho_2)}_{\beta\alpha\beta'\alpha'}
\; .
\eea
Analogous relations hold for all higher density matrices.

We now consider a Hamiltonian of the general form
\be
H=\sum_{\alpha\alpha'}t_{\alpha\alpha'}a_{\alpha}^\dagger a_{\alpha'}
+\frac{1}{2}\sum_{\alpha\beta\alpha'\beta'}{\langle}\alpha\beta|v|
\alpha'\beta'{\rangle}
a_{\alpha}^\dagger a_{\beta}^\dagger a_{\alpha'} a_{\beta'} \; ,
\ee
with a pure two-body interaction $v$.
{}From the hermeticity of the Hamiltonian we get
\be
t_{\alpha\alpha'}=t_{\alpha'\alpha}^* \; , \qquad
{\langle}\alpha\beta|v|\alpha'\beta'{\rangle} \; = \;
{\langle}\alpha'\beta'|v|\alpha\beta{\rangle}^* \;.
\label{hermiterelationen}
\ee
The time evolution of an arbitrary operator $O$ with no explicit time
dependence is given by the Heisenberg equation
\be
i \partial_t O = [ O , H ] \;,
\label{heisenberggleichung}
\ee
where $\partial_t$ denotes the total time derivative.
Then the time derivative of the operator product
\be
a_{\alpha_1'}^\dagger...a_{\alpha_n'}^\dagger a_{\alpha_n}...a_{\alpha_1}=
a_{\alpha_1'}^\dagger...a_{\alpha_n'}^\dagger a_{\alpha_1}...a_{\alpha_n}=
\prod_{i=1}^n a_{\alpha_i'}^\dagger \prod_{k=1}^n a_{\alpha_k} \; ,
\label{operatorprodukt}
\ee
which defines the n-body density matrix (\ref{dmdefinition}) via the
expectation value, reads:
\bea
\lefteqn{ i\partial_t \left[ \left( \prod_{i=1}^n a_{\alpha_i'}^\dagger
\right) \left( \prod_{k=1}^n a_{\alpha_k} \right) \right] } \nonumber \\
& & = \sum_{j=1}^n \sum_\lambda \left\{
t_{\alpha_j \lambda} \left( \prod_{i=1}^n a_{\alpha_i'}^\dagger \right)
\left( \prod_{k\not=j=1}^n a_{\alpha_k} \right) a_\lambda
-t_{\lambda \alpha_j'} \left( \prod_{i\not=j=1}^n
a_{\alpha_j'}^\dagger \right) a_\lambda^\dagger
\left( \prod_{k=1}^n a_{\alpha_k} \right) \right\} \nonumber \\
& & + \frac{1}{2} \sum_{j=1}^n \sum_{k=1}^{j-1} \sum_{\lambda_1 \lambda_2}
{\langle}\alpha_j \alpha_k |v| \lambda_1 \lambda_2{\rangle}_S
\left( \prod_{i=1}^n a_{\alpha_i'}^\dagger \right)
\left( \prod_{l\not=j,k=1}^n a_{\alpha_l} \right)
a_{\lambda_1} a_{\lambda_2} \nonumber \\
& & - \frac{1}{2} \sum_{j=1}^n \sum_{k=j+1}^n \sum_{\lambda_1 \lambda_2}
{\langle}\lambda_1 \lambda_2 |v| \alpha_j' \alpha_k'{\rangle}_S
a_{\lambda_1}^\dagger a_{\lambda_2}^\dagger
\left( \prod_{i\not=j,k=1}^n a_{\alpha_i'}^\dagger \right)
\left( \prod_{l=1}^n a_{\alpha_l} \right) \nonumber \\
& & + \frac{1}{2} \sum_{j=1}^n \sum_{\lambda_1 \lambda_2 \lambda_3}
\left\{ {\langle}\alpha_j \lambda_1 |v| \lambda_2 \lambda_3{\rangle}_S
\left( \prod_{i=1}^n a_{\alpha_i'}^\dagger \right) a_{\lambda_1}^\dagger
\left( \prod_{k\not=j=1}^n a_{\alpha_k} \right)
a_{\lambda_2} a_{\lambda_3} \right. \nonumber \\
& & \left. -{\langle}\lambda_1 \lambda_2 |v| \lambda_3 \alpha_j'{\rangle}_S
\left( \prod_{i\not=j=1}^n a_{\alpha_i'}^\dagger \right)
a_{\lambda_1}^\dagger a_{\lambda_2}^\dagger
\left( \prod_{k=1}^n a_{\alpha_k} \right) a_{\lambda_3} \right\}
\label{bbgkyoperator}
\eea
with the symmetrized two-body matrix elements
\bea
\langle \alpha_j \lambda_1 |v| \lambda_2 \lambda_3 \rangle_S
=\langle \alpha_j \lambda_1 |v| \lambda_2 \lambda_3 \rangle
+ \langle \alpha_j \lambda_1 |v| \lambda_3 \lambda_2 \rangle \; .
\eea
In the middle two terms one can use either $\sum_{k=1}^{j-1}$
or $\sum_{k=j+1}^n$, because the
expressions in the sums are symmetric in $i$ and $j$.
The equations of the BBGKY density matrix hierarchy now follow by
taking the expectation value on both sides of
(\ref{bbgkyoperator})\footnote{One can therefore regard the BBGKY hierarchy
as an operator identity.}.
With (\ref{dmdefinition}) we obtain:
\bea
\lefteqn{ i\partial_t (\rho_n)_{\alpha_1 ... \alpha_n \alpha_1' ...
\alpha_n'}
= \sum_{j=1}^n \sum_{\lambda} \left\{ t_{\alpha_j \lambda}
(\rho_n)_{\alpha_1 ... \alpha_{j-1} \lambda \alpha_{j+1} ... \alpha_n
\alpha_1' ... \alpha_n' } \right. } \nonumber \\
& & \left. -t_{\lambda \alpha_j'} (\rho_n)_{\alpha_1 ... \alpha_n
\alpha_1' ... \alpha_{j-1}' \lambda \alpha_{j+1}' ... \alpha_n' }
\right\} \nonumber \\
& & + \frac{1}{2} \sum_{j=1}^n \sum_{k=1}^{j-1} \sum_{\lambda_1 \lambda_2}
\left\{ {\langle}\alpha_j \alpha_k |v| \lambda_1 \lambda_2 {\rangle}_S
(\rho_n)_{\alpha_1 ... \alpha_{k-1} \lambda_1 \alpha_{k+1} ...
\alpha_{j-1} \lambda_2 \alpha_{j+1} ... \alpha_n \alpha_1' ... \alpha_n'}
\right. \nonumber \\
& & \left. - {\langle} \lambda_1 \lambda_2 |v| \alpha_j' \alpha_k'
{\rangle}_S
(\rho_n)_{\alpha_1 ... \alpha_n \alpha_1' ... \alpha_{k-1}' \lambda_1
\alpha_{k+1}' ... \alpha_{j-1}' \lambda_2 \alpha_{j+1}' ... \alpha_n'}
\right\}  \nonumber \\
& & + \frac{1}{2} \sum_{j=1}^n \sum_{\lambda_1 \lambda_2 \lambda_3}
\left\{ {\langle} \alpha_j \lambda_1 |v| \lambda_2 \lambda_3 {\rangle}_S
(\rho_{n+1})_{\alpha_1 ... \alpha_{j-1} \lambda_2 \lambda_3 \alpha_{j+1} ...
\alpha_n \alpha_1' ... \alpha_n' \lambda_1 } \right. \nonumber \\
& & \left. -{\langle} \lambda_1 \lambda_2 |v| \lambda_3 \alpha_j' {\rangle}_S
(\rho_{n+1})_{ \alpha_1 ... \alpha_n \lambda_3 \alpha_1' ... \alpha_{j-1}'
\lambda_1 \lambda_2 \alpha_{j+1}' ... \alpha_n'} \right\} \; .
\label{bbgkyhierarchie}
\eea
The time evolution of the density matrices consequently is given by a
coupled system of equations of first order in time, where the
equation of motion for $\rho_n$ couples to $\rho_{n+1}$. For practical
purposes the hierarchy (\ref{bbgkyhierarchie}) has to be truncated.

For the special cases of $\rho_1$ and $\rho_2$ -- which we will examine
furtheron -- the equations read explicitly:
\bea
\lefteqn{ i\partial_t (\rho_1)_{\alpha \alpha'} = \sum_\lambda \left\{
t_{\alpha \lambda} (\rho_1)_{\lambda \alpha'} -
t_{\lambda \alpha'} (\rho_1)_{\alpha \lambda} \right\} } \nonumber \\
& & + \frac{1}{2} \sum_{\lambda_1 \lambda_2 \lambda_3} \left\{
{\langle}\alpha \lambda_1 |v| \lambda_2 \lambda_3{\rangle}_S
(\rho_2)_{\lambda_2 \lambda_3 \alpha' \lambda_1}
- {\langle}\lambda_1 \lambda_2 |v| \alpha' \lambda_3 {\rangle}_S
(\rho_2)_{\alpha \lambda_3 \lambda_1 \lambda_2} \right\}
\label{bbgkyeinteilchen}
\eea
and
\bea
\lefteqn{ i\partial_t (\rho_2)_{\alpha \beta \alpha' \beta'} =
\sum_{\lambda} \left\{ t_{\alpha \lambda}
(\rho_2)_{\lambda \beta \alpha' \beta'} + t_{\beta \lambda}
(\rho_2)_{\alpha \lambda \alpha' \beta'}
- t_{\lambda \alpha'} (\rho_2)_{\alpha \beta \lambda \beta'}
- t_{\lambda \beta'} (\rho_2)_{\alpha \beta \alpha' \lambda} \right\} }
\nonumber \\
\nonumber \\
& & + \frac{1}{2} \sum_{\lambda_1 \lambda_2} \left\{
{\langle} \alpha \beta |v| \lambda_1 \lambda_2 {\rangle}_S
(\rho_2)_{\lambda_1 \lambda_2 \alpha' \beta'}
- {\langle} \lambda_1 \lambda_2 |v| \alpha' \beta' {\rangle}_S
(\rho_2)_{\alpha \beta \lambda_1 \lambda_2} \right\} \nonumber \\
\nonumber \\
& & + \frac{1}{2} \sum_{\lambda_1 \lambda_2 \lambda_3} \left\{
{\langle} \alpha \lambda_1 |v| \lambda_2 \lambda_3{\rangle}_S
(\rho_3)_{\beta \lambda_2 \lambda_3 \alpha' \beta' \lambda_1}
+ {\langle} \beta \lambda_1 |v| \lambda_2 \lambda_3 {\rangle}_S
(\rho_3)_{\alpha \lambda_2 \lambda_3 \alpha' \beta' \lambda_1} \right.
\nonumber \\
& & \left. - {\langle} \lambda_1 \lambda_2 |v| \lambda_3 \alpha'{\rangle}_S
(\rho_3)_{\alpha \beta \lambda_3 \beta' \lambda_1 \lambda_2}
- {\langle} \lambda_1 \lambda_2 |v| \lambda_3 \beta'{\rangle}_S
(\rho_3)_{\alpha \beta \lambda_3 \alpha' \lambda_1 \lambda_2} \right\}
\; .
\label{bbgkyzweiteilchen}
\eea

A suitable truncation scheme for (\ref{bbgkyhierarchie}) is based
on the cluster decomposition of Green functions, i.e.
on the fact that any n-point Green function can be decomposed into a sum
of products of connected Green functions of equal or lower order.
The truncation then is performed by
neglecting all connected Green functions higher than a certain order, i.e.
in our case all connected Green functions of higher order than the four-point
function (or two-body correlation function) \cite{23}.
This strategy is based on the assumption that
the connected (correlated) parts of the corresponding Green functions
become increasingly unimportant at higher order \cite{6, 23}.

A reduction to the two-body level requires the equations
of motion for $\rho_1$ (\ref{bbgkyeinteilchen})
and $\rho_2$ (\ref{bbgkyzweiteilchen}), which contain all
density matrices up to
$\rho_3$. Thus we need the cluster expansions of $\rho_1$, $\rho_2$
and $\rho_3$ up to the two-body correlation function.
With $(\rho_1)_{\alpha \alpha'}=\rho_{\alpha \alpha'}$ and $c_n$ denoting the
n-body correlation function we obtain:
\be
(\rho_1)_{\alpha \alpha'}=\rho_{\alpha \alpha'} \; ,
\ee
\bea
\lefteqn{(\rho_2)_{\alpha \beta \alpha' \beta'} =
\rho_{\alpha \alpha'} \rho_{\beta \beta'} + \rho_{\alpha \beta'}
\rho_{\beta \alpha'} + (c_2)_{\alpha \beta \alpha' \beta'} } \nonumber \\
& & =(\rho_{20})_{\alpha \beta \alpha' \beta'}
+(c_2)_{\alpha \beta \alpha' \beta'} \; ,
\eea
\bea
\lefteqn{(\rho_3)_{\alpha \beta \gamma \alpha' \beta' \gamma'} =
\rho_{\alpha \alpha'} \rho_{\beta \beta'} \rho_{\gamma \gamma'} +
\rho_{\alpha \alpha'} \rho_{\beta \gamma'} \rho_{\gamma \beta'} +
\rho_{\alpha \beta'} \rho_{\beta \alpha'} \rho_{\gamma \gamma'} }
\nonumber \\
& & + \rho_{\alpha \beta'} \rho_{\beta \gamma'} \rho_{\gamma \alpha'} +
\rho_{\alpha \gamma'} \rho_{\beta \beta'} \rho_{\gamma \alpha'} +
\rho_{\alpha \gamma'} \rho_{\beta \alpha'} \rho_{\gamma \beta'}
\nonumber \\
& & + \rho_{\alpha \alpha'} (c_2)_{\beta \gamma \beta' \gamma'} +
\rho_{\alpha \beta'} (c_2)_{\beta \gamma \alpha' \gamma'} +
\rho_{\alpha \gamma'} (c_2)_{\beta \gamma \beta' \alpha'}
\nonumber \\
& & + \rho_{\beta \beta'} (c_2)_{\alpha \gamma \alpha' \gamma'} +
\rho_{\beta \alpha'} (c_2)_{\alpha \gamma \beta' \gamma'} +
\rho_{\beta \gamma'} (c_2)_{\alpha \gamma \alpha' \beta'}
\nonumber \\
& & + \rho_{\gamma \gamma'} (c_2)_{\alpha \beta \alpha' \beta'} +
\rho_{\gamma \alpha'} (c_2)_{\alpha \beta \beta' \gamma'} +
\rho_{\gamma \beta'} (c_2)_{\alpha \beta \alpha' \gamma'}
\nonumber \\
& & + (c_3)_{\alpha \beta \gamma \alpha' \beta' \gamma'}
\nonumber \\
& & =(\rho_{30})_{\alpha \beta \gamma \alpha' \beta' \gamma'}
\nonumber \\
& & +(1 + {\cal P}_{\alpha \beta} + {\cal P}_{\alpha \gamma})
(1 + {\cal P}_{\alpha' \beta'} + {\cal P}_{\alpha' \gamma'})
\rho_{\alpha \alpha'} (c_2)_{\beta \gamma \beta' \gamma'}
\nonumber \\
& & + (c_3)_{\alpha \beta \gamma \alpha' \beta' \gamma'} \; ,
\label{clusterexpansion}
\eea
where ${\cal P}_{\alpha \beta}$ is the permutation operator interchanging
the indices $\alpha$ and $\beta$;
\be
(\rho_{20})_{\alpha \beta \alpha' \beta'} =
(1 + {\cal P}_{\alpha \beta}) \rho_{\alpha \alpha'} \rho_{\beta \beta'}
\ee
and
\be
(\rho_{30})_{\alpha \beta \gamma \alpha' \beta' \gamma'} =
(1 + {\cal P}_{\alpha \beta} + {\cal P}_{\alpha \gamma} )
(1 + {\cal P}_{\beta \gamma})
\rho_{\alpha \alpha'} \rho_{\beta \beta'} \rho_{\gamma \gamma'}
\ee
are the uncorrelated parts of the two- and three-body density matrices.

The explicit expressions for
the cluster expansions can be derived from the generating functionals
of full and connected Green functions \cite{24}.

In our present case we make the
additional assumption, that operator products with different numbers of
creation and annihilation operators, i.e. operators not conserving particle
number, have vanishing expectation values, e.g.
\be
{\langle} a_{\alpha}^\dagger a_{\beta}^\dagger{\rangle} \; =
\; {\langle} a_{\alpha} a_{\beta}{\rangle}=0 \; .
\label{partnunonconsexample}
\ee

Neglecting $c_3$ in (\ref{clusterexpansion}) and inserting the
truncated cluster expansions into the equations of motion
(\ref{bbgkyeinteilchen}) and (\ref{bbgkyzweiteilchen}),
we arrive at a closed system of coupled nonlinear differential equations for
$\rho$ and $c_2$. In order to compactify these lengthy equations
(in close analogy to fermionic correlation dynamics \cite{3, 4})
we introduce the following abbreviations:
\be
U_{\alpha \alpha'}=\sum_{\lambda_1 \lambda_2} \rho_{\lambda_1 \lambda_2}
{\langle}\alpha \lambda_2 |v| \alpha' \lambda_1{\rangle}_S
\; \; \; \; {\rm (mean-field)}
\label{meanfielddefinition}
\ee
\be
h_{\alpha \alpha'}=t_{\alpha \alpha'} + U_{\alpha \alpha'}
\; \; \; \; {\rm (mean-field-Hamiltonian)}
\label{meanfieldHamiltoniandefinition}
\ee
\be
Q_{\alpha \beta \alpha' \beta'}^= = \delta_{\alpha \alpha'}
\delta_{\beta \beta'} + \rho_{\alpha \alpha'} \delta_{\beta \beta'}
+ \rho_{\beta \beta'} \delta_{\alpha \alpha'}
\; \; \; \; {\rm (vertical \; \; Bose-factor)}
\label{qparalleldefinition}
\ee
\be
Q_{\alpha \beta \alpha' \beta'}^\perp = \rho_{\beta \alpha'}
\delta_{\alpha \beta'} - \rho_{\alpha \beta'} \delta_{\beta \alpha'}
\; \; \; \; {\rm (horizontal \; \; Bose-factor)} \; .
\label{qsenkrechtdefinition}
\ee
Since no higher correlation functions are included,
$c_2$ is simply denoted by $c$ furtheron.
The compactified equations of motion for $\rho$ and $c$ then read:
\bea
\label{kompakteeinteilchengleichung}
\eqnoinc
\lefteqn{ i\partial_t \rho_{\alpha \alpha'} = \sum_{\lambda} \left(
h_{\alpha \lambda} \rho_{\lambda \alpha'}
- h_{\lambda \alpha'} \rho_{\alpha \lambda} \right)  }
\subeqno \alabel{onemeanfieldpart} \\
\nonumber \\
& & + \sum_{\lambda_1 \lambda_2 \lambda_3} \left(
{\langle}\alpha \lambda_1 |v| \lambda_2 \lambda_3{\rangle}
c_{\lambda_2 \lambda_3 \alpha' \lambda_1}
-{\langle}\lambda_1 \lambda_2 |v| \alpha' \lambda_3{\rangle}
c_{\alpha \lambda_3 \lambda_1 \lambda_2} \right) \; ,
\subeqno \alabel{onecollpart} \\
\nonumber \subeqres
\eea
and
\bea
\label{kompaktezweiteilchengleichung}
\eqnoinc
\lefteqn{ i \partial_t c_{\alpha \beta \alpha' \beta'} =
\sum_\lambda \left( h_{\alpha \lambda} c_{\lambda \beta \alpha' \beta'}
+ h_{\beta \lambda} c_{\alpha \lambda \alpha' \beta'}
- h_{\lambda \alpha'} c_{\alpha \beta \lambda \beta'}
-h_{\lambda \beta'} c_{\alpha \beta \alpha' \lambda} \right) }
\subeqno \alabel{meanfieldpart} \\
\nonumber \\
& & + \sum_{\lambda_1 \lambda_2 \lambda_3 \lambda_4}
{\langle}\lambda_3 \lambda_4 |v| \lambda_1 \lambda_2 {\rangle}
\left\{ Q_{\alpha \beta \lambda_3 \lambda_4}^=
(\rho_{20})_{\lambda_1 \lambda_2 \alpha' \beta'}
 - Q_{\lambda_1 \lambda_2 \alpha' \beta'}^=
(\rho_{20})_{\alpha \beta \lambda_3 \lambda_4} \right\}
\subeqno \alabel{bornpart} \\
\nonumber \\
& & + \sum_{\lambda_1 \lambda_2 \lambda_3 \lambda_4}
{\langle}\lambda_3 \lambda_4 |v| \lambda_1 \lambda_2 {\rangle}
\left\{ Q_{\alpha \beta \lambda_3 \lambda_4}^=
c_{\lambda_1 \lambda_2 \alpha' \beta'}
 - Q_{\lambda_1 \lambda_2 \alpha' \beta'}^=
c_{\alpha \beta \lambda_3 \lambda_4} \right\}
\subeqno \alabel{gmatpart} \\
\nonumber \\
& & + \left( 1 + {\cal P}_{\alpha \beta} \right)
\left( 1 + {\cal P}_{\alpha' \beta'} \right)
\sum_{\lambda_1 \lambda_2 \lambda_3 \lambda_4}
{\langle} \lambda_3 \lambda_4 |v| \lambda_1 \lambda_2{\rangle}_S
Q_{\alpha \lambda_1 \alpha' \lambda_3}^\perp
c_{\lambda_2 \beta \lambda_4 \beta'} \; .
\subeqno \alabel{rpapart} \\
\nonumber \subeqres
\eea
The approximation of the bosonic BBGKY density matrix hierarchy given by
(\ref{kompakteeinteilchengleichung}) and
(\ref{kompaktezweiteilchengleichung}) will be denoted as NCBCD
({\bf N}umber-{\bf C}onserving {\bf B}oson {\bf C}orrelation {\bf D}ynamics)
furtheron. The equations conserve e.g. particle number
${\langle}N{\rangle}=\sum_\alpha \rho_{\alpha \alpha}$
and total energy ${\langle}H{\rangle}$
for a Hamiltonian with no explicit time dependence.

We note in passing that the compactified bosonic NCBCD equations
are formally very similar to the compactified NQCD equations for the
nonrelativistic description of fermionic nuclear systems \cite{4}
with Bose-factors instead of Pauli-blocking-factors and
symmetrized instead of antisymmetrized matrix elements of the two-body
interaction and the two-body density matrix.

The NCBCD equations guarantee a dynamical, nonperturbative resummation
of ring- and ladder-diagrams in both vertical and horizontal direction
and of the corresponding mixed diagrams without double-counting.
Topologically this corresponds to considering parquet-like diagrams.
\section{The diagrammatical structure of the equations of motion}
\label{diagramanalysis}
In this Sect. we investigate the topological structure of the NCBCD
equations of motion for the pionic model specified in Appendix
\ref{specification} and give a graphical representation of the terms
appearing in the equations in order to illustrate the classes of
diagrams that are resummed.

The interaction $v$ (see Appendix \ref{specification}) is split up in the
t-, u- and s-channel part:
\be
{\langle}\alpha \beta |v|\alpha' \beta'{\rangle} \; = \;
{\langle}\alpha \beta |v_{t,u}|\alpha' \beta'{\rangle}
+{\langle}\alpha \beta |v_s|\alpha' \beta'{\rangle} \; .
\ee
In order to allow for a unique identification, we now label the individual
terms in the equations of motion:
\eqnoinc
\bea
\lefteqn{
i\partial_t \rho_{\alpha \alpha'} =
\sum_\lambda \left( t_{\alpha \lambda} \rho_{\lambda \alpha'}
- t_{\lambda \alpha'} \rho_{\alpha \lambda} \right) } \subeqno
\alabel{onekin} \\
\nonumber \\
& & + \sum_\lambda \left( U_{\alpha \lambda} \rho_{\lambda \alpha'}
- U_{\lambda \alpha'} \rho_{\alpha \lambda} \right) \subeqno
\alabel{onemean} \\
\nonumber \\
& & + \sum_{\lambda_1 \lambda_2 \lambda_3} \left(
{\langle}\alpha \lambda_1 |v| \lambda_2 \lambda_3{\rangle}
c_{\lambda_2 \lambda_3 \alpha' \lambda_1}
-{\langle}\lambda_1 \lambda_2 |v| \alpha' \lambda_3{\rangle}
c_{\alpha \lambda_3 \lambda_1 \lambda_2} \right) \, , \subeqno
\alabel{onecoll} \\
\subeqres
\nonumber
\eea
\eqnoinc
\bea
\lefteqn{
i \partial_t c_{\alpha \beta \alpha' \beta'} = \sum_\lambda \left(
t_{\alpha \lambda} c_{\lambda \beta \alpha' \beta'}
+ t_{\beta \lambda} c_{\alpha \lambda \alpha' \beta'}
- t_{\lambda \alpha'} c_{\alpha \beta \lambda \beta'}
- t_{\lambda \beta'} c_{\alpha \beta \alpha' \lambda} \right) }
\subeqno \alabel{twokin} \\
\nonumber \\
& & + \sum_\lambda \left( U_{\alpha \lambda} c_{\lambda \beta \alpha' \beta'}
+ U_{\beta \lambda} c_{\alpha \lambda \alpha' \beta'}
- U_{\lambda \alpha'} c_{\alpha \beta \lambda \beta'}
- U_{\lambda \beta'} c_{\alpha \beta \alpha' \lambda} \right)
\subeqno \alabel{twomean} \\
\nonumber \\
& & + \sum_{\lambda_1 \lambda_2 \lambda_3 \lambda_4}
{\langle}\lambda_3 \lambda_4 |v_{t,u}|\lambda_1 \lambda_2{\rangle}
\left( Q^=_{\alpha \beta \lambda_3 \lambda_4}
(\rho_{20})_{\lambda_1 \lambda_2 \alpha' \beta'}
 - Q^=_{\lambda_1 \lambda_2 \alpha' \beta'}
(\rho_{20})_{\alpha \beta \lambda_3 \lambda_4} \right)
\subeqno \alabel{twobornt} \\
\nonumber \\
& & + \sum_{\lambda_1 \lambda_2 \lambda_3 \lambda_4}
{\langle}\lambda_3 \lambda_4 |v_s|\lambda_1 \lambda_2{\rangle}
\left( Q^=_{\alpha \beta \lambda_3 \lambda_4}
(\rho_{20})_{\lambda_1 \lambda_2 \alpha' \beta'}
 - Q^=_{\lambda_1 \lambda_2 \alpha' \beta'}
(\rho_{20})_{\alpha \beta \lambda_3 \lambda_4} \right)
\subeqno \alabel{twoborns} \\
\nonumber \\
& & + \sum_{\lambda_1 \lambda_2 \lambda_3 \lambda_4}
{\langle}\lambda_3 \lambda_4 |v_{t,u}|\lambda_1 \lambda_2{\rangle}
\left( Q^=_{\alpha \beta \lambda_3 \lambda_4}
c_{\lambda_1 \lambda_2 \alpha' \beta'}
- Q^=_{\lambda_1 \lambda_2 \alpha' \beta'}
c_{\alpha \beta \lambda_3 \lambda_4} \right)
\subeqno \alabel{twogmatt} \\
\nonumber \\
& & + \sum_{\lambda_1 \lambda_2 \lambda_3 \lambda_4}
{\langle}\lambda_3 \lambda_4 |v_s|\lambda_1 \lambda_2{\rangle}
\left( Q^=_{\alpha \beta \lambda_3 \lambda_4}
c_{\lambda_1 \lambda_2 \alpha' \beta'}
 - Q^=_{\lambda_1 \lambda_2 \alpha' \beta'}
c_{\alpha \beta \lambda_3 \lambda_4} \right)
\subeqno \alabel{twogmats} \\
\nonumber \\
& & + \left( 1 + {\cal P}_{\alpha \beta} \right)
\left( 1 + {\cal P}_{\alpha' \beta'} \right)
\sum_{\lambda_1 \lambda_2 \lambda_3 \lambda_4}
{\langle}\lambda_3 \lambda_4 |v_{t,u}| \lambda_1 \lambda_2{\rangle}_S
Q^{\perp}_{\alpha \lambda_1 \alpha' \lambda_3}
c_{\lambda_2 \beta \lambda_4 \beta'} \subeqno \alabel{tworpat} \\
\nonumber \\
& & + \left( 1 + {\cal P}_{\alpha \beta} \right)
\left( 1 + {\cal P}_{\alpha' \beta'} \right)
\sum_{\lambda_1 \lambda_2 \lambda_3 \lambda_4}
{\langle}\lambda_3 \lambda_4 |v_s| \lambda_1 \lambda_2{\rangle}_S
Q^{\perp}_{\alpha \lambda_1 \alpha' \lambda_3}
c_{\lambda_2 \beta \lambda_4 \beta'} \subeqno \alabel{tworpas} \; .\\
\nonumber
\subeqres
\eea
\subsection{The one-body equation}
\label{einterme}
The term (\ref{onekin}) accounts for the free propagation of the one-body
density matrix.
In general all terms are of hermitean structure, i.e. for each
process there is the conjugate process with the incoming and outgoing
indices interchanged.

The term (\ref{onemean}) accounts for the influence of the unrenormalized
mean-field on the propagation of the one-body density matrix,
(\ref{onekin}) and (\ref{onemean}) together then are equivalent to the
TDHF approximation.

In order to investigate the structure of (\ref{onemean}), we first
examine the part with positive overall sign:
\eqnoinc
\bea
\lefteqn{
\sum_\lambda U_{\alpha \lambda} \rho_{\lambda \alpha'} = } \nonumber \\
& & \sum_{\lambda_1 \lambda_2 \lambda_3} \rho_{\lambda_1 \lambda_2}
{\langle}\alpha \lambda_2 |v_{t,u}|\lambda_3 \lambda_1{\rangle}
\rho_{\lambda_3 \alpha'}
\subeqno \alabel{uhartt} \\
\nonumber \\
& & + \sum_{\lambda_1 \lambda_2 \lambda_3} \rho_{\lambda_1 \lambda_2}
{\langle}\alpha \lambda_2 |v_{t,u}|\lambda_1 \lambda_3{\rangle}
\rho_{\lambda_3 \alpha'}  \subeqno \alabel{ufockt} \\
\nonumber \\
& & + \sum_{\lambda_1 \lambda_2 \lambda_3}  \rho_{\lambda_1 \lambda_2}
{\langle}\alpha \lambda_2 |v_s|\lambda_3 \lambda_1{\rangle}
\rho_{\lambda_3 \alpha'}  \subeqno \alabel{uharts} \\
\nonumber \\
& & + \sum_{\lambda_1 \lambda_2 \lambda_3} \rho_{\lambda_1 \lambda_2}
{\langle}\alpha \lambda_2 |v_s|\lambda_1 \lambda_3{\rangle}
\rho_{\lambda_3 \alpha'} \subeqno \alabel{ufocks} \; . \\
\nonumber
\subeqres
\eea
We now use the following convention:
In all diagrams of this Sect. solid lines stand for the link between
two equal indices or for external indices. They describe the propagation of
the pions participating in the process.
In the graphical representation of terms containing a $\sigma$-propagator
we use a horizontal dashed line for a spacelike and a vertical
dashed line for a timelike $\sigma$.
At this point it is useful to note, that in momentum space $v_{t,u}$ is
proportional to a $\sigma$-propagator depending on the difference of two
on-shell four momenta while $v_s$ is proportional to a $\sigma$-propagator
containing the sum of two on-shell four momenta, where for
$v_s$ (up to the formfactor) the factor multiplying the propagator is only
half of that for $v_{t,u}$\footnote{The difference of two on-shell
four momenta of particles of equal rest mass is always spacelike, the sum
is always timelike.} (see Appendix \ref{specification}).

The contributions (\ref{uhartt}) and (\ref{uharts}) are depicted in
fig. \ref{onehart} and correspond to the Hartree approximation. The terms
(\ref{ufockt}) and (\ref{ufocks}) are depicted in fig. \ref{onefock} and
correspond to the exchange terms in the Hartree-Fock approximation.
The graphical representation of the conjugate terms is obtained
by interchanging the incoming and outgoing external indices.

The coupling of the equation of motion for the one-body density matrix
to the two-body correlation function is given by (\ref{onecoll}).
In fig. \ref{bildonecoll} we show only one of the two conjugate
processes. Both conjugate processes together have a gain-loss structure
as it is known e.g. from collision terms in transport equations \cite{1}.
However, (\ref{onecoll}) cannot directly be identified with a collision
term, since it also renormalizes the mean-field.

In order to present the last point more explicitly, it is useful to
introduce a G-matrix and a selfenergy within the formalism of equal-time
density matrices.
The G-matrix $G_{\alpha \beta \alpha' \beta'}$ is defined by
\bea
\sum_{\lambda_1 \lambda_2} G_{\alpha \beta \lambda_1 \lambda_2}
(\rho_{20})_{\lambda_1 \lambda_2 \alpha' \beta'} =
\sum_{\lambda_1 \lambda_2} {\langle}\alpha \beta |v| \lambda_1
\lambda_2{\rangle}
(\rho_2)_{\lambda_1 \lambda_2 \alpha' \beta'} \; .
\label{gmatrixdef}
\eea
Inserting (\ref{gmatrixdef}) into the equation of motion for $\rho$ leads to
\bea
i \partial_t \rho_{\alpha \alpha'} = \sum_\lambda
\left( t_{\alpha \lambda} \rho_{\lambda \alpha'}
- t_{\lambda \alpha'} \rho_{\alpha \lambda} \right)
+ \sum_{\lambda_1 \lambda_2 \lambda_3}
\left( G^S_{\alpha \lambda_3 \lambda_1 \lambda_2}
\rho_{\lambda_1 \alpha'} \rho_{\lambda_2 \lambda_3}
-G^S_{\lambda_1 \lambda_2 \alpha' \lambda_3}
\rho_{\alpha \lambda_1} \rho_{\lambda_3 \lambda_2} \right)
\label{onebodygmat}
\eea
with $G^S_{\alpha \beta \alpha' \beta'} = G_{\alpha \beta \alpha' \beta'}
+ G_{\alpha \beta \beta' \alpha'}$.
The G-matrix terms in (\ref{onebodygmat}) contain all
corrections to the free propagation, i.e. also the renormalized mean-field.

A further compactification of the one-body equation can be achieved by
introducing the selfenergy according to
\bea
\Sigma_{\alpha \alpha'} = \sum_{\lambda_1 \lambda_2}
\rho_{\lambda_1 \lambda_2}
G^S_{\alpha \lambda_2 \alpha' \lambda_1} \; .
\label{selfendef}
\eea
Inserting (\ref{selfendef}) into (\ref{onebodygmat}) gives
\be
i \partial_t \rho_{\alpha \alpha'} = \sum_\lambda \left\{
\left( t_{\alpha \lambda} + \Sigma_{\alpha \lambda} \right)
\rho_{\lambda \alpha'}
-\left( t_{\lambda \alpha'} + \Sigma_{\lambda \alpha'} \right)
\rho_{\alpha \lambda} \right\} \; .
\label{onebodyselfen}
\ee
The renormalized mean-field can now be obtained as the hermitean part of
$\Sigma_{\alpha \alpha'}$, i.e.
\bea
\lefteqn{
U^{ren}_{\alpha \alpha'}=\left( Re \Sigma \right)_{\alpha \alpha'}
= \frac{1}{2} \left( \Sigma + \Sigma^\dagger \right)_{\alpha \alpha'}
= \frac{1}{2} \Sigma_{\alpha \alpha'}
+ \frac{1}{2} \Sigma^*_{\alpha' \alpha} } \nonumber \\
& & = \frac{1}{2} \sum_{\lambda_1 \lambda_2} \left(
\rho_{\lambda_1 \lambda_2} G^S_{\alpha \lambda_2 \alpha' \lambda_1}
+ \rho^*_{\lambda_1 \lambda_2} G^{S*}_{\alpha' \lambda \alpha \lambda_1}
\right)
= \sum_{\lambda_1 \lambda_2} \rho_{\lambda_1 \lambda_2}
\left( Re G^S \right)_{\alpha \lambda_2 \alpha' \lambda_1} \; ,
\label{hermsigma}
\eea
i.e. as the selfenergy obtained from the hermitean part
of the symmetrized G-matrix.
The renormalized mean-field Hamiltonian can then be defined by
\be
h^{ren}_{\alpha \alpha'} = t_{\alpha \alpha'} + U^{ren}_{\alpha \alpha'} \; .
\label{renormh}
\ee
With the above definitions the equation of motion for the one-body density
matrix reads
\bea
i\partial_t \rho_{\alpha \alpha'}-\sum_\lambda \left\{
h^{ren}_{\alpha \lambda} \rho_{\lambda \alpha'}
-h^{ren}_{\lambda \alpha'} \rho_{\alpha \lambda} \right\} =
\sum_\lambda \left\{ (Im\Sigma)_{\alpha \lambda} \rho_{\lambda \alpha'}
- \rho_{\alpha \lambda} (Im\Sigma)_{\lambda \alpha'}  \right\} \; ,
\label{transport}
\eea
where the term on the right hand side of (\ref{transport})
corresponds to the collision term of a transport theory.
The explicit derivation of a transport equation with a collision term
containing the correct in-medium factors and transport coefficients from
the NCBCD equations can be carried out in complete analogy to the
fermionic case \cite{1, 25}.
\subsection{The mean-field terms in the two-body equation}
\label{twobmefi}
The term (\ref{twokin}) in analogy to the corresponding one in the
one-body equation
accounts for the free propagation of the two-body correlation function.

The mean-field part (\ref{twomean}) of the two-body equation has the
same structure as in (\ref{onemean});
for each of the external indices there is a diagram with a mean-field
loop at the corresponding line.
\subsection{The Born-term}
\label{twobborn}
The expressions (\ref{twobornt}) and (\ref{twoborns}) only contain matrix
elements of the two-body interaction coupled to a Bose-factor $Q^=$ and two
one-body density matrices, but not the two-body correlation function itself;
they describe exactly one elementary interaction process between two
pions and thus correspond to the first Born approximation of scattering
theory, i.e. (\ref{twobornt}) in the t- and u- channel and (\ref{twoborns})
in the s-channel. The two terms together are hence referred to as Born-term.

The scattering process is subject to an in-medium modification by the
Bose-factor in the incoming or (for the conjugate process) in the outgoing
channel. Using the single particle basis that diagonalizes $\rho$, i.e.
\be
\rho_{\alpha \alpha'}=\delta_{\alpha \alpha'} n_\alpha \; ,
\ee
the summations over the two intermediate indices appearing in $Q^=$
in (\ref{twoborns}) and (\ref{twobornt}) break down and we obtain
factors of $1+n_\alpha+n_\beta$ in the incoming channel and
$1+n_\alpha'+n_\beta'$ in the outgoing channel; the Bose-factors thus
cause an enhancement of the contribution of the corresponding process
to the total amplitude proportional to the occupation numbers of the
respective states. As an example, two of the diagrams generated by
(\ref{twobornt}) and (\ref{twoborns}) are represented in
fig. \ref{bildtwoborn}.
\subsection{Dynamical iteration in the vertical direction}
\label{twobvert}
The terms (\ref{twogmatt}) and (\ref{twogmats}) each contain a matrix
element of the two-body correlation function itself
coupled to a matrix element of the two-body potential,
leading to a dynamical iteration of the interaction in
''vertical'' direction (i.e. along the time direction of the diagrams).
The same Bose-factor $Q^=$ as for the Born-term appears in the incoming or
the outgoing channel. The two terms with $Q^=$ in the incoming channel are
graphically represented in fig. \ref{bildtwogmat}.

The left diagram in fig. \ref{bildtwogmat} shows, that the term
(\ref{twogmatt}) leads to a dynamical
resummation of ladder-diagrams in vertical direction, where in
the intermediate states as well as in one of the external channels
there is a Bose-factor $Q^=$.
The right diagram in fig. \ref{bildtwogmat} shows, that the term
(\ref{twogmats}) leads to a dynamical
resummation of ring-diagrams in vertical direction with the same
factors of $Q^=$ appearing as for the ladder diagrams.
The first step of the resummation in each case is given
by the Born-term; (\ref{twogmatt}) and (\ref{twogmats}) together lead to a
mutual dynamical iteration of the ring- and ladder-diagrams
as shown in fig. \ref{bildtwogmat}.

In analogy to the fermionic NQCD case one obtains a Bethe-Goldstone
equation for the G-matrix by neglecting (\ref{tworpat}) and (\ref{tworpas})
and considering the equations in the stationary limit.
The terms (\ref{twogmatt}) and (\ref{twogmats}) are therefore in the
following denoted as G-matrix-terms.

In order to simplify the notation, the single particle basis is now chosen
to diagonalize the unrenormalized mean-field Hamiltonian, i.e.
\be
h_{\alpha \alpha'}=\delta_{\alpha \alpha'} \epsilon_\alpha \; .
\ee
One can then rewrite the equation of motion for $c$ in the limit
described above as
\bea
\lefteqn{
\left( \omega-\epsilon_\alpha-\epsilon_\beta \right)
c_{\alpha \beta \alpha' \beta'}
-\left( \omega-\epsilon_{\alpha'}-\epsilon_{\beta'} \right)
c_{\alpha \beta \alpha' \beta'} } \nonumber \\
& &  = \sum_{\lambda_1 \lambda_2 \lambda_3 \lambda_4}
{\langle}\lambda_3 \lambda_4 |v| \lambda_1 \lambda_2{\rangle}
 \left\{ Q^=_{\alpha \beta \lambda_3 \lambda_4}
(\rho_2)_{\lambda_1 \lambda_2 \alpha' \beta'}
-Q^=_{\lambda_1 \lambda_2 \alpha' \beta'}
(\rho_2)_{\alpha \beta \lambda_3 \lambda_4} \right\} \; . \nonumber \\
\eea
This equation is fulfilled for ($\eta \rightarrow 0^+$), if
\be
c_{\alpha \beta \alpha' \beta'}=
\frac{1}{\omega - \epsilon_\alpha - \epsilon_\beta + i \eta}
\sum_{\lambda_1 \lambda_2 \lambda_3 \lambda_4}
Q^=_{\alpha \beta \lambda_3 \lambda_4}
G_{\lambda_3 \lambda_4 \lambda_1 \lambda_2}
(\rho_{20})_{\lambda_1 \lambda_2 \alpha' \beta'}
\label{invert}
\ee
and the conjugate equation hold, where at the right hand side the
definition of the G-matrix has been inserted.
Matrix-multiplication of (\ref{invert}) with the matrix element of the
two-body potential and eliminating $c$ on the left hand side via
$c=\rho_2-\rho_{20}$ leads to
\bea
\lefteqn{
\sum_{\gamma_1 \gamma_2} G_{\alpha \beta \gamma_1 \gamma_2}
(\rho_{20})_{\gamma_1 \gamma_2 \alpha' \beta'}
= \sum_{\gamma_1 \gamma_2}
{\langle}\alpha \beta |v|\gamma_1 \gamma_2{\rangle}
(\rho_{20})_{\gamma_1 \gamma_2 \alpha' \beta'}  } \nonumber \\
& & + \sum_{\gamma_1 \gamma_2 \lambda_1 \lambda_2 \lambda_3 \lambda_4}
{\langle}\alpha \beta |v|\gamma_1 \gamma_2{\rangle}
\frac{Q^=_{\gamma_1 \gamma_2 \lambda_3 \lambda_4}}{\omega-
\epsilon_{\gamma_1} - \epsilon_{\gamma_2} + i \eta }
G_{\lambda_3 \lambda_4 \lambda_1 \lambda_2}
(\rho_{20})_{\lambda_1 \lambda_2 \alpha' \beta'} \nonumber \\
\eea
for an arbitrary $\rho_{20}$, i.e. we have
\bea
G_{\alpha \beta \alpha' \beta'} = \; {\langle}\alpha \beta |v| \alpha' \beta'
{\rangle}
+ \sum_{\lambda_1 \lambda_2 \lambda_3 \lambda_4}
{\langle}\alpha \beta |v| \lambda_1 \lambda_2{\rangle}
\frac{Q^=_{\lambda_1 \lambda_2 \lambda_3 \lambda_4}}{\omega-
\epsilon_\alpha - \epsilon_\beta + i \eta }
G_{\lambda_3 \lambda_4 \alpha' \beta'} \; .
\label{bethegoldstone}
\eea
Equation (\ref{bethegoldstone}) is a Bethe-Goldstone-equation for the
G-matrix which guarantees the resummation of ring- and ladder-diagrams
in vertical direction.
\subsection{Dynamical iteration in the horizontal direction}
\label{twobhori}
The terms (\ref{tworpat}) and (\ref{tworpas}) each also contain a
two-body correlation function coupled to a matrix element of the
two-body potential, which leads to a dynamical resummation of diagrams.
The topological structure of the terms, graphically represented in
figs. \ref{bildtworpat} and \ref{bildtworpas}, in contrast to
(\ref{twogmatt}) and (\ref{twogmats}) leads to a resummation in
''horizontal'' direction, where the factor $Q^\perp$ appears in the
intermediate states and at a pair of an incoming and an outgoing index.

In the single particle basis with
$\rho_{\alpha \alpha'}=\delta_{\alpha \alpha'} n_\alpha$,
as for the terms accounting for the resummation in vertical direction,
the summation over the intermediate indices
breaks down and one remains with a factor of $n_\alpha' - n_\alpha$.
In our notation each term containing one horizontal factor $Q^\perp$
comprises a process together with its conjugate process.

The direct term in (\ref{tworpat}), with the indices in the matrix element
of the two-body potential not interchanged, is shown in
fig. \ref{bildtworpat} at the top. It leads to a dynamical resummation
of ring-diagrams in horizontal direction.
The exchange term in (\ref{tworpat}) accounts for the dynamical resummation
of ladder-diagrams in horizontal direction.

In (\ref{tworpas}) the direct term and the exchange term are equal;
together they account for the resummation of ladder-diagrams in
horizontal direction.

All terms of the NCBCD approximation together lead to a dynamical resummation
and mutual iteration of ring- and ladder-diagrams in vertical and
horizontal direction on the two-body level, i.e. a parquet-like resummation
\cite{26, 27} which also renormalizes the one-body equation of motion.
\section{Numerical studies for pionic systems}
\label{numerical}
\subsection{Comparison of partial resummations}
\label{partial}
In the following we consider a pion gas as described in Appendix
\ref{specification} confined to a box of
sidelengths $L_x=L_y=L_z=10{\rm fm}$.
In order to reduce the numerical effort we assume the pion gas to be
in an isospin-symmetric configuration and thereby eliminate the internal
isospin degrees of freedom as described in Appendix \ref{eqomiso}.
As a single particle basis for our
numerical simulations we choose standing waves, which vanish at the box
boundaries. The calculations are performed using the 11
(in the noninteracting case) energetically lowest basis elements.

Since the NCBCD equations only describe the propagation of the density
matrices in time, we first have to solve the problem of finding proper
initial conditions for the system at some given starting time.
In the bosonic case there is no direct
access to the manifold of stationary solutions of the NCBCD equations
and, furthermore, it is e.g. not clear, what further constraints --
besides the above mentioned symmetry properties -- these
solutions have to fulfill in order to remain physical.
In \cite{28} it was shown for the case of an analytically solvable model
(Lipkin-model), that only requiring the density matrices to be positively
definite and to satisfy the trace theorems for fermionic systems with
good particle number leads to a sensible solution within the framework of
the NQCD formalism. However, in our case the trace theorems do not hold.

The most obvious starting point for the initial condition is given by a
stationary self-consistent Hartree-Fock solution. In order to generate
this solution for a given temperature and particle number, the one-body
density matrix is occupied with a Bose-Einstein distribution with respect
to the eigenstates of the mean-field Hamiltonian, where the chemical
potential is adjusted in order to give the right particle number.

In fig. \ref{bildene} the Hartree-Fock energy levels and occupation numbers
are shown for the example of systems with 60 and 90 pions for various
temperatures, where the occupation numbers are given for only one
isospin quantum number; the total occupation numbers are obtained by
multiplication with a factor of 3. Due to the spatial symmetries of the
system, some of the higher lying levels are degenerate.
As one expects, the error due to the finite number of basis elements
reduces in going to higher densities and lower thermal excitations.

After providing a first approximation for the initial condition of the
pion gas, we now investigate how the system evolves in time using the
complete NCBCD equations or certain limiting cases for the partial
resummation of diagrams on the two-body level.
We aim at deciding, if for an approximately correct description of two-body
quantities in the pionic many-body system one really needs to use the
complete, numerically very involved NCBCD equations.

The limiting case, in which only the resummation of ring- and
ladder-diagrams in vertical direction is taken into account, is obtained
from the complete equations (\ref{kompakteeinteilchengleichung}) and
(\ref{kompaktezweiteilchengleichung}) by neglecting (\ref{rpapart})
and corresponds to time-dependent G-Matrix theory (TDGMT). It will be
denoted by {\bf vertical approximation} in the following.

The limiting case, which only takes into account the resummation of ring-
and ladder-diagrams in horizontal direction, is obtained from
(\ref{kompakteeinteilchengleichung}) and
(\ref{kompaktezweiteilchengleichung}) by neglecting (\ref{gmatpart}) and
will be denoted by {\bf horizontal approximation} furtheron.
In the fermionic case the approximation generated from the NQCD equations
in an analogous way can be identified with a two-body RPA theory
\cite{2, 5}.

Using the isospin channel quantities defined in Appendix \ref{eqomiso},
we get the expression for the total energy of the system:
\bea
\label{isototen}
\eqnoinc
\lefteqn{
E=\;{\langle}H{\rangle}\;=3 \sum_{\alpha \alpha'} t_{\alpha' \alpha}
\rho_{\alpha \alpha'} }
\subeqno \alabel{isokinen} \\
& & + \frac{1}{2} \sum_{\alpha \alpha' \beta \beta'}
{\langle}\alpha' \beta'|v_{tu}|\alpha \beta{\rangle}
\left( 9 \rho_{\alpha \alpha'} \rho_{\beta \beta'}
+ 3 \rho_{\alpha \beta'} \rho_{\beta \alpha'} \right)
\nonumber \\
& & + \frac{1}{2} \sum_{\alpha \alpha' \beta \beta'}
{\langle}\alpha' \beta'|v_s| \alpha \beta{\rangle}
 3 (\rho_{20})_{\alpha \beta \alpha' \beta'}
\subeqno \alabel{isomeanen}  \\
 & & + \frac{1}{2} \sum_{\alpha \alpha' \beta \beta'}
 {\langle}\alpha' \beta'|v_{tu}|\alpha \beta{\rangle}
 \left( 3 c^{stu}_{\alpha \beta \alpha' \beta'}
+ 6 c^{t1}_{\alpha \beta \alpha' \beta'} \right)
\nonumber \\
& & + \frac{1}{2} \sum_{\alpha \alpha' \beta \beta'}
{\langle}\alpha' \beta'|v_s| \alpha  \beta{\rangle}
\left( 3 c^{stu}_{\alpha \beta \alpha' \beta'}
+ 6 c^s_{\alpha \beta \alpha' \beta'} \right) \; .
\subeqno \alabel{isocorren} \\
\nonumber
\subeqres
\eea
The kinetic energy is given by (\ref{isokinen}), the mean-field energy by
(\ref{isomeanen}) and the correlation energy by (\ref{isocorren});
the total energy, which is conserved in time, is the sum of these
three contributions.

Even though in the investigation for fermionic systems in \cite{5}
the vertical as well as
the horizontal approximation showed distinct deviations from the full
NQCD approximation, the correlation energies still had about
the same order of magnitude as in the complete correlation dynamics approach.
For the pionic systems under investigation here this is no longer true.
While at lower densities the 3 approximations do not differ
significantly, at higher densities the horizontal approximation yields
absolute correlation energies which are drastically higher
than those extracted from the NCBCD calculations.

In order to illustrate this behaviour,
we show in the upper part of fig. \ref{bildavcorre}
the correlation energies (averaged over time)
for the 3 approximations as a function of the pion number at a temperature
of 100 MeV. The system in each case was propagated up to the time
$t=20 \; {\rm fm}/c$. One clearly sees the dramatic overestimation of
correlation energy by the horizontal approximation with increasing density.
In the lower part of fig. \ref{bildavcorre} we have left
out the horizontal approximation in order to allow for a better comparison
between NCBCD and the vertical approximation. The vertical approximation
always generates a correlation energy, which is about a factor of 2
too small as compared to the NCBCD correlation energy.

This demonstrates the importance of mixed diagrams consisting of
contributions from the vertical and the horizontal approximation and
of the interference of the different classes of diagrams.
We are thus lead to the
conclusion, that only the full NCBCD approach is able to yield
adequate results for two-body quantities in the pionic system
under consideration.
Especially we infer, that a description of two-body dynamics by a
Bethe-Goldstone approach or a time-dependent G-Matrix theory is not
appropriate.
For all further investigations we, therefore,
use the complete NCBCD approach.
\subsection{Dynamical generation of a correlated state}
\label{generation}
In this Sect. we will demonstrate, how one can
dynamically generate an approximately stationary physical state for the
NCBCD equations, which is fully correlated on the two-body level.

A first attempt for the solution of this problem consists in
initializing the Hartree-Fock solution and propagating the system in
time until all correlations have built up and the system has reached
a quasi-stationary state. However, in this state all
density matrix elements will
show more or less stable oscillations about an equilibrium value and not
necessarily approach a stationary state asymptotically,
as we have checked numerically.
Furthermore, this approach has the additional problem, that the proper
energy of the correlated state cannot be reached, since the NCBCD
approximation conserves total energy.

For the reasons mentioned above we introduce a different method by
modifying the NCBCD equations of motion in the following way:
The terms (\ref{onemeanfieldpart}) and (\ref{meanfieldpart}),
which describe the free and mean-field
propagation of $\rho$ and $c$, remain unchanged.
In all other terms, i.e. in (\ref{onecollpart}),(\ref{bornpart}),
(\ref{gmatpart}) and (\ref{rpapart}), the matrix elements of the
two-body potential ${\langle}\alpha \beta |v| \alpha' \beta'{\rangle}$
are multiplied by a time-dependent dimensionless factor $g(t)$.
This alteration of the equations of motion, although at first sight
appearing to be arbitrary, does not lead to an unphysical propagation of
the system, since it is equivalent to an alteration of the Hamiltonian
according to $H \rightarrow H'(t)$, where
\bea
\lefteqn{
H'(t)=\sum_{\alpha \alpha'} \left\{ t_{\alpha \alpha'}
+ \left( 1-g(t) \right) \sum_{\lambda_1 \lambda_2} \rho_{\lambda_1 \lambda_2}
{\langle}\alpha \lambda_2 |v|\alpha' \lambda_1{\rangle}_S \right\}
a_\alpha^\dagger a_{\alpha'} } \nonumber \\
& & + g(t) \sum_{\alpha \beta \alpha' \beta'}
{\langle}\alpha \beta |v| \alpha' \beta'{\rangle}
a_\alpha^\dagger a_\beta^\dagger a_{\alpha'} a_{\beta'}.
\label{modiham}
\eea
If $g(t)=0$, one recovers the
TDHF approximation; if $g(t)=1$ one recovers the full NCBCD approximation
with respect to $H$.

By initializing a stationary Hartree-Fock solution at time $t=0$
and propagating the system using the NCBCD equations induced by
(\ref{modiham}) up to the time $t=t_f$, where $g(0)=0$ and $g(t_f)=1$,
we continuously switch on the residual interaction on the two-body level.
The total energy is not conserved
during this process, thus enabling the system to energetically approach
the correlated equilibrium configuration.

Numerical experiments using different forms of $g(t)$ have shown
that a linear function requires the smallest amount of computation time
in order to guarantee the convergence of the method.
We, therefore, choose $g(t)=\lambda t$ and hence $t_f=1/\lambda$.

In fig. \ref{bildconv} the change $\Delta E$ of the
total energy reached at the end of the
process is plotted versus $t_f$ for 30 pions in the volume of
$10 \; {\rm fm}^3$.
The picture clearly shows, that there is a convergence
as $t_f \to \infty$ (adiabatic limit); i.e. if the parameter
$g(t)$ changes slowly, the response of the system is fast enough
to generate an equilibrium configuration with respect to $H'(t)$ at
any given time $t$. The last statement implies, that for $t_f \to \infty$
a convergence of the trajectory of the system in the overall configuration
space has to occur. This is indeed the case, as can e.g. be seen
from fig. \ref{bildcorrconv}, where the correlation energy with respect to
$H$ during the process is depicted as a function of $t/t_f$ for various
values of $t_f$. Furthermore, fig. \ref{bildcorrconv} illustrates,
that the oscillations caused by switching on the residual interaction
too fast progressively vanish as one goes to higher values of $t_f$,
which clearly indicates that indeed in the limit $t_f \to \infty$
the system equilibrates at each time during the process.

The change in total energy during the adiabatic process is not mainly
caused by a buildup of correlation energy, but rather by a change in
kinetic and mean-field energy by about the same order of magnitude,
since the generation of two-body correlations causes a redistribution at
the one-body level as well.
This can be seen from fig. \ref{bildabcd}, where the changes in the
single contributions to the total energy and the change of the total energy
with respect to the Hartree-Fock solution are plotted as a function of $t_f$.
Furthermore, we note that for all densities and temperatures
considered in this work, the total energy is lowered when the residual
interaction is adiabatically switched on.
\subsection{Two-body observables in coordinate space}
\label{coordinate}
We now turn to the investigation of the two-body density matrix resulting
from the adiabatic method described above. For this purpose the residual
interaction in each case is switched on with $t_f=160 \; {\rm fm}/c$,
and the values for $\rho$ and $c$ reached at $t=t_f$ are used for
the evaluation of the observables of interest.

The two-body density matrix $\rho_2$ in coordinate space is given by
\bea
\rho_2 (\vec{r}_1, \vec{r}_2; \vec{r'}_1, \vec{r'}_2)=
\sum_{\alpha \beta \alpha' \beta'} (\rho_2)_{\alpha \beta \alpha' \beta'}
\psi_{\alpha}(\vec{r}_1) \psi_{\beta}(\vec{r}_2)
\psi_{\alpha'}^*(\vec{r'}_1) \psi_{\beta'}^*(\vec{r'}_2)
\eea
and the uncorrelated part of $\rho_2$, i.e. $\rho_{20}$,
and the two-body correlation function $c$ by
\bea
\rho_{20} (\vec{r}_1, \vec{r}_2; \vec{r'}_1, \vec{r'}_2)=
\sum_{\alpha \beta \alpha' \beta'} (\rho_{20})_{\alpha \beta \alpha' \beta'}
\; \psi_{\alpha}(\vec{r}_1) \psi_{\beta}(\vec{r}_2)
\psi_{\alpha'}^*(\vec{r'}_1) \psi_{\beta'}^*(\vec{r'}_2)
\eea
and
\bea
c (\vec{r}_1, \vec{r}_2; \vec{r'}_1, \vec{r'}_2)=
\sum_{\alpha \beta \alpha' \beta'} c_{\alpha \beta \alpha' \beta'}
\; \psi_{\alpha}(\vec{r}_1) \psi_{\beta}(\vec{r}_2)
\psi_{\alpha'}^*(\vec{r'}_1) \psi_{\beta'}^*(\vec{r'}_2) \; ,
\eea
with
\bea
\rho_2 (\vec{r}_1, \vec{r}_2; \vec{r'}_1, \vec{r'}_2) =
\rho_{20} (\vec{r}_1, \vec{r}_2; \vec{r'}_1, \vec{r'}_2) +
c (\vec{r}_1, \vec{r}_2; \vec{r'}_1, \vec{r'}_2)
\eea
and
\bea
\rho_{20} (\vec{r}_1, \vec{r}_2; \vec{r'}_1, \vec{r'}_2) =
\rho (\vec{r}_1; \vec{r'}_1) \rho (\vec{r}_2; \vec{r'}_2) +
\rho (\vec{r}_1; \vec{r'}_2) \rho (\vec{r}_2; \vec{r'}_1) \; .
\eea

The probability distribution for a simultaneous measurement of a particle
at position $\vec{r}_1$ and a particle at $\vec{r}_2$ is given by
$\rho_2(\vec{r}_1, \vec{r}_2; \vec{r}_1, \vec{r}_2)$.
In fig. \ref{bild50150x}
$\rho(x,x;x,x)=\int dy dz \rho_2 (\vec{r}, \vec{r}; \vec{r}, \vec{r})$
with $\vec{r}=(x,y,z)$ is plotted as a function of $x$ for the temperatures
50 and 150 MeV at various pion densities (solid line).
In addition, the same quantity is shown without the correlated part
of $\rho_2$, i.e. only for $\rho_{20}$ (dashed line), and for the
classical two-body distribution
\bea
\rho_{2}^{class} (\vec{r}_1, \vec{r}_2; \vec{r'}_1, \vec{r'}_2) =
\rho (\vec{r}_1; \vec{r'}_1) \rho (\vec{r}_2; \vec{r'}_2) \; ,
\eea
i.e. neglecting Bose statistics (dotted line). At the boundaries
of the box all distributions vanish due to the boundary conditions
imposed on the single particle basis elements.
The inclusion of Bose statistics leads to an enhancement
of the distribution obtained with $\rho_{20}$ by a factor of 2 with
respect to the distribution obtained
with $\rho_2^{class}$. Including the two-body correlation
function leads to a further increase in the probability of finding two
particles at the same place, which is due to the fact that the
$\sigma$-interaction used is purely attractive.

We now investigate the strength of the enhancement of
$\rho_2(\vec{r},\vec{r};\vec{r},\vec{r})$
relative to \\ $\rho(\vec{r};\vec{r}) \rho(\vec{r};\vec{r})$
as a function of temperature and density.
Naively one might expect, that with increasing density
the system will be more correlated and therefore show a stronger
enhancement. However, this is not true.
In fig. \ref{bild50150x} one can already observe,
that at equal temperature an increase in density does not necessarily
imply an increase in the relative enhancement of the
two-body probability. While at $T$=50 MeV the enhancement decreases
with increasing density over the whole density range considered, at
$T$=150 MeV there is a maximum in enhancement between 45 and 60 pions
in $(10 \; {\rm fm})^3$.

In order to illustrate this point more clearly the quantity
\bea
\frac{\rho_2}{\rho\rho}=
\frac{\int d^3r \rho_2 (\vec{r},\vec{r};\vec{r},\vec{r})}{\int
d^3 r \rho(\vec{r};\vec{r}) \rho(\vec{r};\vec{r}) }
\eea
is plotted in fig. \ref{bildvett} as a function of the pion density
for various temperatures.
For higher temperatures there are distinct maxima of the
relative enhancement at certain densities. For all temperatures the
relative enhancement then decreases again for higher densities.
This behaviour is an indication of the nonperturbative nature of
n-point correlation dynamics (the lowest order of which in our case is given
by the Hartree-Fock approximation), since a denser and therefore more
strongly interacting system does not imply a worse convergence of the
cluster expansion.

The next point to consider is the probability distribution for
simultaneously finding two particles at different places in coordinate space.
The question is, how this two-body density is altered by the inclusion
of the two-body correlation function as a function of the
relative distance of the two particles.

In fig. \ref{bild3d50m10200m30} a two-dimensional cut
through the six-dimensional quantities
\bea
\frac{\rho_{20}(\vec{r}_1, \vec{r}_2; \vec{r}_1, \vec{r}_2)}{\rho(\vec{r}_1;
\vec{r}_1) \rho(\vec{r}_2; \vec{r}_2)} \; ; \quad
\frac{\rho_2(\vec{r}_1, \vec{r}_2; \vec{r}_1, \vec{r}_2)}{\rho(\vec{r}_1;
\vec{r}_1) \rho(\vec{r}_2; \vec{r}_2)}
\label{6dim}
\eea
at $y_1=y_2=z_1=z_2=5 \; {\rm fm}$ (i.e. in the center of the box of
sidelength $10 \; {\rm fm}$) is depicted as a contour-plot as a function
of $x_1$ and $x_2$ for two parameter pairs of density and temperature; in
the left part of the figure only Bose-statistics are taken into account.
The inclusion of the two-body correlation function (right part) leads to an
increased gradient with respect to the relative distance, i.e. at smaller
distances the two-body density is increased and at larger distances
decreased.

In fig. \ref{bild50150m} we show the same quantity as a function of $x$
with $x_1=x$ and $x_2=L-x$, i.e. along a cut from the center to the lower
right corner in the previous figure, for 50 and 150 MeV temperature and
various densities, where $L=10 \; {\rm fm}$ is the sidelength of the box;
the two-body densities evaluated with $\rho_2$ (solid lines)
and with $\rho_{20}$ (dashed lines) are directly compared.

Due to Bose-statistics, the curve for $\rho_{20}$ always approaches 2 as
$x \to 0$. Again the increased gradient with respect to the relative
distance caused by the inclusion of the two-body correlation function can
be seen, which is most pronounced for those parameters, that also show
the strongest enhancement of probability at relative distance 0, i.e. at
the maxima in fig. \ref{bildvett}.
\subsection{Relative strength of two-body correlations}
\label{relative}
In the last Sect. we discovered, that for fixed temperature the relative
enhancement of the probability for simultaneously detecting two particles
at the same place in coordinate space exhibits a maximum at a certain
density. This behaviour of the pionic many-body system will now be
analyzed more closely.
In this respect, we introduce the quantity
\bea
\frac{c}{\rho_2}=
\frac{\sum_{\alpha \beta \alpha' \beta'} |c_{\alpha \beta \alpha' \beta'}|}{
\sum_{\alpha \beta \alpha' \beta'}|(\rho_2)_{\alpha \beta \alpha' \beta'}|}
\; ,
\label{correlationstrength}
\eea
which will be denoted as ''correlation strength'' in the following.
While in the two-body density in coordinate space only the isospin
channel correlation functions $c^{stu}$ contribute (since we are looking
at pairs of identical pions), all possible isospin configurations enter
into the evaluation of the correlation strength (\ref{correlationstrength}).
Also all off-diagonal elements of the two-body density matrix are
considered without having the possibility of mutually cancelling out
each other. Although (\ref{correlationstrength}) is not invariant
under unitary transformations of the single particle basis,
it is still a useful measure of the relative importance of two-body
correlations in the system.

The curves in fig. \ref{bildfpeaks} were generated by adiabatically
switching on the residual interaction as described in Sect. \ref{generation}
using $t_f=160 \;{\rm fm}/c$ and plotting the correlation strength reached
at $t=t_f$ versus density for various temperatures.
Again there are distinct maxima as in the case of the relative
enhancement of the two-body density in coordinate space.
If we compare fig. \ref{bildfpeaks} to fig. \ref{bildvett},
we observe a very good agreement of the shape of the curves, implying
that the observed result is not an artifact of the way we defined our
measure for the relative importance of two-body correlations.

As one approaches the maxima, the stationary Hartree-Fock solution
used for the initialization at $t=0$ becomes increasingly useless.
It is therefore tempting to assume, that the convergence of the cluster
expansion in general might also become worse;
an assumption which would have to be checked by explicitly taking into
account higher order than the two-body correlation functions.
However, within the framework of the NCBCD approximation it is not
possible to verify this assumption.

A possible interpretation of the observed maximum is
that the system runs through a second order phase transition at a certain
critical density (for fixed temperature), which corresponds to a critical
chemical potential in the parameter space of the theory.
In such a scenario one might expect the cluster expansion to break down as
one approaches the critical region.
Moving away from the critical region the mean-field
(Hartree-Fock) solution progressively becomes better
and the convergence of the cluster expansion improves \cite{24}.

A problem we want to point out in this context is that in a finite
system, as considered in this work, one cannot draw any definite
conclusions about the real order of a phase transition, since in such
systems usually also first order phase transitions are smeared out and
appear as second order phase transitions \cite{29}.

Finally, we explicitly demonstrate that not only a higher density, but also
a stronger coupling does not automatically imply stronger correlations in
the pionic system. For that purpose we show in fig. \ref{bilddouble}
the correlation strength obtained by adiabatically switching on the residual
interaction plotted versus density for 100 MeV temperature for the coupling
given by (\ref{parameters}) (solid line) and for a coupling twice
as large (dashed line). The correlation strength for the higher value of
the coupling constant is distinctly smaller than that for the lower one.
This again clearly proves the nonperturbative nature of the cluster
expansion.
\section{Summary}
\label{summary}
In this work we have derived the NCBCD
approximation for the nonperturbative, dynamical description of particle
number conserving bosonic many-body systems including two-body correlations.
This approach leads to a dynamical resummation and mutual iteration of
ring- and ladder-diagrams in vertical and horizontal direction on the
two-body level, i.e. a parquet-like resummation \cite{26, 27}, which also
renormalizes the one-body equation of motion.

For our numerical simulations we assumed a pion gas confined to a box
and a single particle basis of standing waves. The $\pi\pi$ interaction
used was derived from a covariant Lagrangian in a three-dimensional
reduction scheme and fitted to $\pi\pi$ phase shifts. Within this model
we showed, that a restriction of the equations to a resummation in
horizontal direction leads to a dramatic overestimation of correlation
energy and a restriction to a resummation in vertical direction
leads to an underestimation of correlation energy by about a factor of 2.
Thus an adequate description of dynamical two-body
correlations in pionic systems requires the use of the complete NCBCD
equations; e.g. a description within the TDGMT approximation is
inappropriate.

We, furthermore, showed a way to obtain
a correlated, stationary and physical configuration by adiabatically
switching on the residual interactions on the two-body level starting
from a stationary Hartree-Fock solution. For this configuration
we investigated two-body probability distributions in coordinate space
and obtained an increased probability for the simultaneous detection of
two pions at the same place and a decreased probability for the simultaneous
detection of two pions at large distances caused by the inclusion of the
two-body correlation function.
The two-body correlations in coordinate space as well as
the correlation strength (defined in Sect. \ref{relative}) showed a
distinct maximum in the relative strength of two-body correlations as
compared to the disconnected parts of the two-body density matrix
at a certain critical density at fixed temperature.
This might be interpreted as
a signature for a phase transition at a certain critical density.
However, in the investigated
temperature region of $T \le 200 \; {\rm MeV}$ this maximum in each case was
located at a density below $0.09/{\rm fm}^3$, which is much smaller than
those estimated in ultrarelativistic nucleus-nucleus collisions \cite{10}.

Furthermore we found, that at higher densities the two-body correlations
progressively decrease in relative importance, which implies the
possibility of an adequate description of the system on the one-body level
in such cases as ultrarelativistic heavy-ion collisions. We thus conclude
that the conventional Hanbury Brown-Twiss (HBT) analysis of pion sources
from $\pi-\pi$ correlations in such reactions appears justified.
\section*{Appendix}
\begin{appendix}
\section{Specification of the model Hamiltonian}
\label{specification}
In order to describe an interacting pion system we consider the Lagrangian
\be
{\cal L} = \frac{1}{2} \left( \partial_\mu \vec{\Phi} \right)
\left( \partial^\mu \vec{\Phi} \right) - \frac{1}{2} m^2 \vec{\Phi}^2
+ \frac{1}{2} \left( \partial_\mu \sigma \right)
\left( \partial^\mu \sigma \right) - \frac{1}{2} M^2 \sigma^2
- g \sigma \vec{\Phi}^2 \; ,
\label{lagrangedichte}
\ee
where $\vec{\Phi}=\left( \Phi_1 , \Phi_2 , \Phi_3 \right)$ is the
pion field of mass $m$ and $\sigma$ is the scalar, isoscalar
field of mass $M$ mediating the interaction between the pions.

{}From the Lagrangian (\ref{lagrangedichte}) we get the coupled equations
of motion for the classical fields or equivalently for the quantized field
operators of the system:
\be
\partial^\mu \partial_\mu \Phi_i + m^2 \Phi_i = - 2 g \sigma \Phi_i
\label{phieom}
\ee
and
\be
\partial^\mu \partial_\mu \sigma + M^2 \sigma = - g \vec{\Phi}^2 \; .
\label{sigmaeom}
\ee
In order to obtain a two-body interaction for the pions, we integrate
out the $\sigma$-field.
Formally inverting (\ref{sigmaeom}) we get
\bea
\tilde{ \sigma } (k) =
\frac{1}{k^\mu k_\mu - M^2 + i \epsilon} g \int d^4 x \; e^{i k^\mu x_\mu}
\vec{\Phi}^2 (x) \; .
\label{sigmaschlange}
\eea
With $\tilde{G}_\sigma (k) = \frac{1}{k^\mu k_\mu - M^2 + i \epsilon}$
and $ G_\sigma (x) = \int \frac{d^4 k}{(2\pi)^4}
e^{- i k^\mu x_\mu } \tilde{G}_\sigma (k)$ we have
\bea
\tilde{\sigma} (k) = g \tilde{G}_\sigma (k)
\int d^4 x' \;  e^{i k^\mu x_\mu'} \vec{\Phi}^2 (x')
\Rightarrow \sigma (x) = g \int d^4 x' G_\sigma (x-x')
\vec{\Phi}^2 (x') \; .
\label{sigmaausintegriert}
\eea
Insertion of (\ref{sigmaausintegriert}) into (\ref{phieom}) leads to
\be
\left( \partial^\mu \partial_\mu + m^2 \right) \Phi_i (x) =
-2 g^2 \Phi_i (x) \int d^4 x' G_\sigma (x-x') \vec{\Phi}^2 (x') \; ,
\label{phiausintegriert}
\ee
i.e. the $\sigma$-field has been eliminated from the equations of motion.

We aim at a Hamiltonian, which yields (\ref{phiausintegriert}) via the
Heisenberg equations for
$\Phi_i$ and its canonically conjugate field momentum $\Pi_i$.

This implies, that we have to carry out a three-dimensional reduction of the
four-dimensional integration in (\ref{phiausintegriert}).
As a first step we neglect the zeroth component of the four momentum
$k^\mu$ in the denominator of $\tilde{G}_\sigma (k)$, i.e.
\be
\tilde{G}_\sigma (k) \approx \tilde{G}_\sigma (\vec{k}) =
\frac{1}{-\vec{k}^2 - M^2}
\label{instantanerpropagatork}
\ee
and
\be
G_\sigma (\vec{x}) = \int \frac{d^3 k}{(2\pi)^3} e^{i \vec{k} \vec{x}}
\frac{1}{-\vec{k}^2 - M^2} \; .
\label{instantanerpropagatorx}
\ee
The intrinsic flaws of this naive instantaneous approximation will in part
be cured later on. The easiest way to uncover these defects is to use a
mode-expansion of the pion fields and to establish,
which terms are responsible for s-, t- or u-channel scattering.

With (\ref{instantanerpropagatork}) and (\ref{instantanerpropagatorx})
we get
\bea
\lefteqn{ \int d^4 x' G_\sigma (x-x') \vec{\Phi}^2 (x') \approx
\int d^4 x' \int \frac{d^4 k}{2\pi)^4} e^{-i k^\mu (x_\mu - x_\mu')}
\frac{1}{-\vec{k}^2 - M^2} \vec{\Phi}^2 (x') } \nonumber \\
& & = \int d^4 x' \int \frac{d^3 k}{(2\pi)^3} e^{i \vec{k}(\vec{x}
-\vec{x}')} \vec{\Phi}^2 (x') \frac{1}{-\vec{k}^2 - M^2}
\delta (x^0 - {x'}^0 ) \nonumber \\
& & = \int d^3 x' G_\sigma (\vec{x} - \vec{x}') \vec{\Phi}^2 (\vec{x}',x^0)
\; .
\label{instantanesintegral}
\eea

The equation of motion from Hamiltonian dynamics in the
instantaneous approximation (\ref{instantanesintegral}) thus reads:
\be
\left( \partial^\mu \partial_\mu + m^2 \right) \Phi_i (\vec{x}) =
-2 g^2 \Phi_i (\vec{x}) \int d^3 x' G_\sigma (\vec{x}-\vec{x}')
\vec{\Phi}^2 (\vec{x}') \; ,
\label{instantanegleichung}
\ee
where all field operators are taken at the same time $t$.
This is fulfilled for
\bea
\lefteqn{
H = \frac{1}{2} \int d^3 x \left\{ \vec{\Pi}^2 (\vec{x}) +
\left( \nabla \vec{\Phi} (\vec{x}) \right)^2 +
m^2 \vec{\Phi}^2 (\vec{x}) \right\} } \nonumber \\
& & + \frac{1}{2} g^2 \int d^3 x d^3 x' \vec{\Phi}^2 (\vec{x})
G_\sigma (\vec{x} - \vec{x}') \vec{\Phi}^2 (\vec{x}')
\; ,
\label{Hamiltonian1}
\eea
since (\ref{Hamiltonian1}) together with
(\ref{heisenberggleichung}) and the canonical equal-time commutation
relations
\beano
\left[ \Phi_i (\vec{x},t) , \Phi_j (\vec{x}',t) \right] =
\left[ \Pi_i (\vec{x},t) , \Pi_j (\vec{x}',t) \right] = 0 \; ,
\eeano
\bea
\left[ \Phi_i (\vec{x},t) , \Pi_j (\vec{x}',t) \right] =
i \delta_{ij} \delta^{(3)} (\vec{x} - \vec{x}')
\label{kanonkommu}
\eea
generate the equations of motion
\be
\partial_t \Phi_i (\vec{x}) = \Pi_i (\vec{x})
\ee
and
\be
\partial_t \Pi_i (\vec{x}) = \nabla^2 \Phi_i (\vec{x})
- m^2 \Phi_i (\vec{x}) - 2 g^2 \Phi_i (\vec{x})
\int d^3 x' G_\sigma (\vec{x} - \vec{x}') \vec{\Phi}^2 (\vec{x}') \; ,
\ee
equivalent to (\ref{instantanegleichung}).

The mode expansions for the pion field operators now read:
\be
\Phi_i (\vec{x}) = \int \frac{d^3 k}{(2\pi)^\frac{3}{2}}
\frac{ e^{i \vec{k} \vec{x} } }{\sqrt{2 \omega (\vec{k}) }}
\left[ a (\vec{k},i) + a^\dagger (-\vec{k},i) \right]
\label{modent1}
\ee
and
\be
\Pi_i (\vec{x}) = \int \frac{d^3 k}{(2\pi)^\frac{3}{2}}
e^{i \vec{k} \vec{x}} \sqrt{\frac{\omega( \vec{k} )}{2} }
\left[ a (\vec{k},i) - a^\dagger (-\vec{k},i) \right]
\label{modent2}
\ee
with
\be
\omega (\vec{k})= \sqrt{ \vec{k}^2 + m^2 }
\ee
and
\be
\left[ a (\vec{k},i) ,  a^\dagger (\vec{k}',j) \right] =
\delta_{ij} \delta^{(3)} (\vec{k} - \vec{k}') \; ,
\ee
where $i$ denotes the isospin of the corresponding operator with regard to
the cartesian representation of SU(2).
Inserting the mode expansions (\ref{modent1}) and (\ref{modent2}) into
(\ref{Hamiltonian1}) leads to
\beano
\lefteqn{ H = \sum_{\tau} \int d^3 k \; \omega (\vec{k})
a^\dagger (\vec{k},\tau) a (\vec{k},\tau)  }
\nonumber \\
& & + \sum_{\tau \tau'} \int d^3 k_1 d^3 k_2 d^3 k_3 d^3 k_4
\frac{g^2}{8(2\pi)^3} \left( \omega(\vec{k}_1) \omega(\vec{k}_2)
\omega(\vec{k}_3) \omega(\vec{k}_4) \right)^{-\frac{1}{2}}
\nonumber \\
& & \times \delta^{(3)} (\vec{k}_1 + \vec{k}_2 + \vec{k}_3 + \vec{k}_4)
\tilde{G}_\sigma (\vec{k}_1 + \vec{k}_2) \times
\nonumber \\
& & \times \left\{
\underline{a(\vec{k}_1,\tau) a(\vec{k}_2,\tau) a(\vec{k}_3,\tau')
a(\vec{k}_4,\tau') }
+ \underline{a(\vec{k}_1,\tau) a(\vec{k}_2,\tau) a(\vec{k}_3,\tau')
a^\dagger (-\vec{k}_4,\tau') } \right.
\nonumber \\
& & + \underline{ a(\vec{k}_1,\tau) a(\vec{k}_2,\tau)
a^\dagger (-\vec{k}_3,\tau') a(\vec{k}_4,\tau') }
+ a(\vec{k}_1,\tau) a(\vec{k}_2,\tau) a^\dagger (-\vec{k}_3,\tau')
a^\dagger (-\vec{k}_4,\tau')
\nonumber \\
& & + \underline{ a(\vec{k}_1,\tau) a^\dagger (-\vec{k}_2,\tau)
a(\vec{k}_3,\tau') a(\vec{k}_4,\tau') }
+ a(\vec{k}_1,\tau) a^\dagger (-\vec{k}_2,\tau) a(\vec{k}_3,\tau')
a^\dagger (-\vec{k}_4,\tau')
\nonumber \\
& & + a(\vec{k}_1,\tau) a^\dagger (-\vec{k}_2,\tau)
a^\dagger (-\vec{k}_3,\tau') a(\vec{k}_4,\tau')
+ \underline{ a(\vec{k}_1,\tau) a^\dagger (-\vec{k}_2,\tau)
a^\dagger (-\vec{k}_3,\tau') a^\dagger (-\vec{k}_4,\tau') }
\nonumber \\
& & + \underline{ a^\dagger (-\vec{k}_1,\tau) a(\vec{k}_2,\tau)
a(\vec{k}_3,\tau') a(\vec{k}_4,\tau') }
+ a^\dagger (-\vec{k}_1,\tau) a(\vec{k}_2,\tau) a(\vec{k}_3,\tau')
a^\dagger (-\vec{k}_4,\tau')
\nonumber \\
& & + a^\dagger (-\vec{k}_1,\tau) a(\vec{k}_2,\tau)
a^\dagger (-\vec{k}_3,\tau') a(\vec{k}_4,\tau')
+ \underline{ a^\dagger (-\vec{k}_1,\tau) a(\vec{k}_2,\tau)
a^\dagger (-\vec{k}_3,\tau') a^\dagger (-\vec{k}_4,\tau') }
\nonumber \\
& & + a^\dagger (-\vec{k}_1,\tau) a^\dagger (-\vec{k}_2,\tau)
a(\vec{k}_3,\tau') a(\vec{k}_4,\tau')
+ \underline{ a^\dagger (-\vec{k}_1,\tau) a^\dagger (-\vec{k}_2,\tau)
a(\vec{k}_3,\tau') a^\dagger (-\vec{k}_4,\tau') }
\nonumber \\
& & \left. + \underline{ a^\dagger (-\vec{k}_1,\tau) a^\dagger
(-\vec{k}_2,\tau)
a^\dagger (-\vec{k}_3,\tau') a(\vec{k}_4,\tau') }
+ \underline{ a^\dagger (-\vec{k}_1,\tau)
a^\dagger (-\vec{k}_2,\tau) a^\dagger (-\vec{k}_3,\tau')
a^\dagger (-\vec{k}_4,\tau') } \right\} \; ,
\eeano
\bea
\label{modeexpansion}
\eea
where now $\tau$ denotes the isospin quantum number.

The Hamiltonian (\ref{modeexpansion}) still contains terms (underlined)
not commuting with the particle number operator.
In view of Sect. \ref{eqom} we neglect such terms
in our approach, i.e. we only consider elastic scattering of pions
(also including particle-number conserving
s-channel scattering).
This is equivalent to the assumption, that the
-- particle number violating -- operator products
generated by the underlined terms in (\ref{modeexpansion})
(after normal ordering) in the equations of motion for particle number
conserving operator products have vanishing or at least negligible
expectation values, in agreement with assumption (\ref{partnunonconsexample})
of Sect. \ref{eqom}.

Normal ordering the remaining terms in (\ref{modeexpansion}) gives:
\bea
\label{hammi}
\eqnoinc
\lefteqn{ H = \sum_{\tau} \int d^3 k \; \omega (\vec{k})
a^\dagger (\vec{k},\tau) a (\vec{k},\tau)  } \nonumber \\
& & + \sum_{\tau \tau'} \int d^3 k_1 d^3 k_2 d^3 k_3 d^3 k_4
\frac{g^2}{8(2\pi)^3} \left( \omega(\vec{k}_1) \omega(\vec{k}_2)
\omega(\vec{k}_3) \omega(\vec{k}_4) \right)^{-\frac{1}{2}}
\nonumber \\
& & \times \delta^{(3)} (\vec{k}_1 + \vec{k}_2 + \vec{k}_3 + \vec{k}_4)
\tilde{G}_\sigma (\vec{k}_1 + \vec{k}_2)  \nonumber \\
& & \times \left\{ a^\dagger (-\vec{k}_3,\tau') a^\dagger (-\vec{k}_4,\tau')
a(\vec{k}_1,\tau) a(\vec{k}_2,\tau)
+ a^\dagger (-\vec{k}_2,\tau) a^\dagger (-\vec{k}_4,\tau')
a(\vec{k}_1,\tau) a(\vec{k}_3,\tau')  \right. \nonumber \\
& & + a^\dagger (-\vec{k}_2,\tau) a^\dagger (-\vec{k}_3,\tau')
a(\vec{k}_1,\tau) a(\vec{k}_4,\tau')
+ a^\dagger (-\vec{k}_1,\tau) a^\dagger (-\vec{k}_4,\tau')
a(\vec{k}_2,\tau) a(\vec{k}_3,\tau')  \nonumber \\
& & \left. + a^\dagger (-\vec{k}_1,\tau) a^\dagger (-\vec{k}_3,\tau')
a(\vec{k}_2,\tau) a(\vec{k}_4,\tau')
+ a^\dagger (-\vec{k}_1,\tau) a^\dagger (-\vec{k}_2,\tau)
a(\vec{k}_3,\tau') a(\vec{k}_4,\tau') \right\} \nonumber \\
\nonumber \\
& & = \sum_{\tau} \int d^3 k \; \omega (\vec{k})
a^\dagger (\vec{k},\tau) a (\vec{k},\tau)  \nonumber \\
& & + \sum_{\tau \tau'} \int d^3 k_1 d^3 k_2 d^3 k_3 d^3 k_4
\frac{g^2}{8(2\pi)^3} \left( \omega(\vec{k}_1) \omega(\vec{k}_2)
\omega(\vec{k}_3) \omega(\vec{k}_4) \right)^{-\frac{1}{2}}
\delta^{(3)} (\vec{k}_1 + \vec{k}_2 - \vec{k}_3 - \vec{k}_4)
\nonumber \\
& & \times \left\{ 2 \tilde{G}_\sigma (\vec{k}_1 + \vec{k}_2)
a^\dagger (\vec{k}_1,\tau) a^\dagger (\vec{k}_2,\tau)
a(\vec{k}_3,\tau') a(\vec{k}_4,\tau') \right. \subeqno
\alabel{schannel} \\
& & \left. + 4 \tilde{G}_\sigma (\vec{k}_1 - \vec{k}_3)
a^\dagger (\vec{k}_1,\tau) a^\dagger (\vec{k}_2,\tau')
a(\vec{k}_3,\tau) a(\vec{k}_4,\tau') \right\} \subeqno \alabel{tchannel} \; .
\\
\nonumber
\subeqres
\eea

As advertised before, one can now extract the terms corresponding to the
different scattering channels; i.e.
(\ref{schannel}) is responsible for the s-channel scattering of pions,
since the intermediate $\sigma$-propagator depends on the sum of
momenta in the in- or outgoing channel and the (cartesian)
isospin quantum numbers of the two pions in each of these channels have to
be equal.

Term (\ref{tchannel}) is responsible for
t- and u-channel scattering of pions,
since the intermediate $\sigma$-propagator depends on the momentum transfer
from one of the incoming to one of the outgoing particles, while these two
particles have to carry the same isospin quantum number as well as the
remaining two pions.

We are now in the position to improve the three-dimensional reduction scheme.
In the center of momentum-frame of two colliding pions the zeroth component
of the four momentum in the $\sigma$-propagator vanishes for t- or u-channel
scattering, while for s-channel scattering the three momentum in the
$\sigma$-propagator vanishes.
This implies that the naive instantaneous approximation
(\ref{instantanerpropagatork}) is inappropriate for the s-channel
term of the two-body potential.

The situation is now cured by simply placing the zeroth components of
the pion four-momenta on-shell.
The ambiguity, which of the -- up to now equivalent -- three momentum
combinations in the $\sigma$-propagator have to be used,
we resolve by averaging over both possibilities. This method is in close
analogy to the well-known three-dimensional reduction schemes of Gross,
Blankenbecler-Sugar and Thompson \cite{30}, except for the difference,
that with these methods one carries out the three-dimensional reduction of
T- or G-matrix equations, where in addition to the external three momenta
the invariant mass $\sqrt{s}$ of the colliding system is fixed.
In the latter case one has different possibilities of placing the particles
off-shell, such that their four momenta give the correct $\sqrt{s}$.
In a dynamical many-body calculation as in this work, the $\sqrt{s}$
for an elementary two-body scattering process can only be reconstructed
from the three-momenta involved, so that the schemes cited above are not
directly applicable.

Since point-like interaction concepts are inappropriate for hadron
scattering we introduce formfactors for the $\pi\pi\sigma$-vertices
containing momentum cutoffs to regularize the theory.
We use a formfactor of the form
\be
F(k^2)=\frac{2(\Lambda^2 - M^2) \Lambda^2 + M^4}{2(\Lambda^2-k^2)\Lambda^2
+ k^4} \; ,
\label{formfak}
\ee
where $\Lambda$ is the cutoff parameter and $k$ is the four momentum
of the $\sigma$-particle coupling to the vertex.
The formfactor (\ref{formfak}) has the distinct advantage
of leading to a uniform parameterization for spacelike and timelike four
momenta without running into a pole in one of the two domains.

The Hamiltonian we use in our explicit computations then reads:
\bea
\lefteqn{ H = \sum_\tau \int d^3 k \; \omega(\vec{k})
a^\dagger (\vec{k},\tau) a(\vec{k},\tau) } \nonumber \\
& & + \frac{1}{2} \sum_{\tau \tau'} \int d^3 k_1 d^3 k_2 d^3 k_3 d^3 k_4 \;
\delta^{(3)}(\vec{k}_1+\vec{k}_2-\vec{k}_3-\vec{k}_4) \nonumber \\
& & \times \left\{ {\langle}\vec{k}_1 \vec{k}_2 |v_s| \vec{k}_3
\vec{k}_4{\rangle}
a^\dagger (\vec{k}_1,\tau) a^\dagger (\vec{k}_2,\tau)
a(\vec{k}_3,\tau') a(\vec{k}_4,\tau') \right. \nonumber \\
& & \left. + {\langle}\vec{k}_1 \vec{k}_2 |v_{t,u}| \vec{k}_3
\vec{k}_4{\rangle}
a^\dagger (\vec{k}_1,\tau) a^\dagger (\vec{k}_2,\tau')
a(\vec{k}_3,\tau) a(\vec{k}_4,\tau') \right\}
\label{fertigerHamiltonian}
\eea
with
\bea
\lefteqn{ {\langle}\vec{k}_1 \vec{k}_2 |v_s| \vec{k}_3 \vec{k}_4{\rangle}=
\frac{g^2}{4(2\pi)^3} \left( \omega(\vec{k}_1) \omega(\vec{k}_2)
\omega(\vec{k}_3) \omega(\vec{k}_4) \right)^{-\frac{1}{2}} }
\nonumber \\
& & \times \left\{ \frac{\left\{F\left( (\omega(\vec{k}_1)
+\omega(\vec{k}_2))^2
-(\vec{k}_1 + \vec{k}_2)^2 \right) \right\}^2 }{(\omega(\vec{k}_1)
+\omega(\vec{k}_2))^2 - (\vec{k}_1 + \vec{k}_2)^2 - M^2}  \right.
\nonumber \\
& & \left. + \frac{\left\{F\left( (\omega(\vec{k}_3)+\omega(\vec{k}_4))^2
-(\vec{k}_3 + \vec{k}_4)^2 \right) \right\}^2 }{(\omega(\vec{k}_3)
+\omega(\vec{k}_4))^2 - (\vec{k}_3 + \vec{k}_4)^2 - M^2} \right\}
\label{vs}
\eea
and
\bea
\lefteqn{ {\langle}\vec{k}_1 \vec{k}_2 |v_{t,u}| \vec{k}_3
\vec{k}_4{\rangle}=
\frac{g^2}{2(2\pi)^3} \left( \omega(\vec{k}_1) \omega(\vec{k}_2)
\omega(\vec{k}_3) \omega(\vec{k}_4) \right)^{-\frac{1}{2}} }
\nonumber \\
& & \times \left\{ \frac{\left\{F\left( (\omega(\vec{k}_1)
-\omega(\vec{k}_3))^2
-(\vec{k}_1 - \vec{k}_3)^2 \right) \right\}^2 }{(\omega(\vec{k}_1)
-\omega(\vec{k}_3))^2 - (\vec{k}_1 - \vec{k}_3)^2 - M^2}  \right.
\nonumber \\
& & \left. + \frac{\left\{F\left( (\omega(\vec{k}_2)-\omega(\vec{k}_4))^2
-(\vec{k}_2 - \vec{k}_4)^2 \right) \right\}^2 }{(\omega(\vec{k}_2)
-\omega(\vec{k}_4))^2 - (\vec{k}_2 - \vec{k}_4)^2 - M^2} \right\}
\; .
\label{vtu}
\eea

The $\sigma$ can be associated with the resonances in the
($J^{PC}=0^{++}$, $I^{G}=0^+$)-channel ($\sigma$-channel) in
$\pi\pi$-scattering, where the theory described can easily be
generalized to a coupling of the pion field to more than a single
scalar and isoscalar field; we simply obtain the matrix
elements of the two-body potential as a sum over those of the
form (\ref{vs}), (\ref{vtu}) for the
individual $\sigma$-particles $\sigma_i$ of masses $M_i$
with couplings $g_i$ and cutoffs $\Lambda_i$.
The parameters used in this work are taken from \cite{31}\footnote{We use
a slightly different form of the formfactor, which however agrees reasonably
well with their form in the kinematical regime under investigation.},
where additional
particles besides the scalar and isoscalar ones are used to fit scattering
data, but the couplings for the latter are directly deduced from the
widths of the resonances in the $\sigma$-channel by means of a Breit-Wigner
formula. Using only $\sigma$-mesons -- in order to keep the
numerical task within a reasonable range -- should at least guarantee a
realistic description of the short-range attractive part of the
$\pi\pi$-interaction.
The parameters taken from \cite{31} are:
\bea
&M_1&=980 \; {\rm MeV} , \; \Lambda_1=1200 \; {\rm MeV} , \;
g_1=595 \; {\rm MeV} \nonumber \\
&M_2&=1300 \; {\rm MeV} , \; \Lambda_2=1200 \; {\rm MeV} , \;
g_2=1854 \; {\rm MeV} \nonumber \\
&m&=m_\pi=140 \; {\rm MeV} \; .
\label{parameters}
\eea
In view of the complexity of the NCBCD equations we do not claim to
have a fully realistic model of a pion gas, which would require the
inclusion of $\rho$-mesons and kaons in a coupled channel-calculation
\cite{32}-\cite{38} in order to fit the experimentally
measured phase shifts for the higher partial waves \cite{39, 40}.
The interaction should work at moderate relative momenta.
\section{Equations of motion for the isosymmetric \protect \\ pion system}
\label{eqomiso}
Since for a numerical solution of the NCBCD equations
the required computation time increases like $N^6$ and the required
memory space increases like $N^4$, where $N$ denotes the number of
single particle basis states, it is furthermore desirable to eliminate
all internal degrees of freedom from the equations and thereby reduce
$N$ by a factor of 3.
We therefore assume, that the system is in an isospin-symmetric
state and that all density matrices are diagonal in isospin space, which
in the spherical representation of isospin SU(2) implies, that the sum of
the isospin-z-components in the incoming and outgoing channels have to be
equal. As one easily verifies, isosymmetry and isodiagonality are conserved
dynamically; therefore these assumptions do not lead to an inconsistent
theory.

Now we use $\tilde{\alpha}=(\alpha,\tau_\alpha)$, where $\alpha$ denotes the
quantum numbers of the space-time degrees of freedom in an arbitrary
orthonormal single particle basis and $\tau$ denotes
the isospin quantum number in the cartesian representation.
The Hamiltonian of the pionic system can then be written as
\bea
\lefteqn{
H=\sum_{\tilde{\alpha} \tilde{\alpha}'} t_{\tilde{\alpha} \tilde{\alpha}'}
a^\dagger_{\tilde{\alpha}} a_{\tilde{\alpha}'}
+ \frac{1}{2}
\sum_{\tilde{\alpha} \tilde{\beta} \tilde{\alpha}' \tilde{\beta}'}
{\langle}\tilde{\alpha} \tilde{\beta} |v| \tilde{\alpha'}
\tilde{\beta'}{\rangle}
a^\dagger_{\tilde{\alpha}} a^\dagger_{\tilde{\beta}}
a_{\tilde{\alpha}'} a_{\tilde{\beta}'} }
\label{tildeHamiltonian}
\eea
with
\be
t_{\tilde{\alpha} \tilde{\alpha}'} = t_{\alpha \alpha'}
\delta_{\tau_\alpha \tau_{\alpha}'}
\ee
and
\bea
{\langle}\tilde{\alpha} \tilde{\beta} |v| \tilde{\alpha}' \tilde{\beta}'
{\rangle} =
{\langle}\alpha \beta |v_s| \alpha' \beta'{\rangle}
\delta_{\tau_\alpha \tau_\beta}
\delta_{\tau_{\alpha'} \tau_{\beta'}}
+ {\langle}\alpha \beta |v_{t,u}| \alpha' \beta'{\rangle}
\delta_{\tau_\alpha \tau_{\alpha'}} \delta_{\tau_\beta \tau_{\beta'}}
\; .
\eea

Isosymmetry implies, that in the cartesian representation of SU(2)
all density matrices have to be invariant under an overall permutation
of isospin quantum numbers.

For the one-body density matrices we assume
\be
\rho_{\tilde{\alpha} \tilde{\alpha}'}=\rho_{\alpha \alpha'}
\delta_{\tau_\alpha \tau_{\alpha'}} \; .
\ee

For the two-body correlation function
$c_{\tilde{\alpha} \tilde{\beta} \tilde{\alpha'} \tilde{\beta'}}$
there are 4 possible distinct isospin configurations:
\bea
&1.& \tau_\alpha = \tau_\beta = \tau_{\alpha'} = \tau_{\beta'}
\nonumber \\
&2.& \tau_\alpha = \tau_\beta \; , \; \tau_{\alpha'} =\tau_{\beta'}
\; , \; \tau_\alpha \not= \tau_{\alpha'}
\nonumber \\
&3.& \tau_\alpha=\tau_{\alpha'} \; , \; \tau_\beta=\tau_{\beta'}
\; , \; \tau_\alpha \not= \tau_\beta
\nonumber \\
&4.& \tau_\alpha=\tau_{\beta'} \; , \; \tau_\beta=\tau_{\alpha'}
\; , \; \tau_\alpha \not= \tau_\beta \; .
\label{possibilities}
\eea
All other configurations, which do not contain two pairs of equal
isospin indices, violate the diagonality of the two-body density matrix
in isospin space; the corresponding correlation functions are thus
assumed to vanish.

Configuration 1. corresponds to the situation, where s-, t- and u-channel
scattering contribute to the Born amplitude for the elementary process;
configuration 2. corresponds to the situation, where only s-channel
scattering contributes and configurations 3. and 4. reflect the
situation, where only t-channel scattering contributes.

We can therefore decompose the two-body correlation function in the
following way:
\bea
\lefteqn{
c_{\tilde{\alpha} \tilde{\beta} \tilde{\alpha}' \tilde{\beta}'} =
\delta_{\tau_\alpha \tau_\beta} \delta_{\tau_{\alpha'} \tau_{\beta'}}
\delta_{\tau_\alpha \tau_{\alpha'}}
c^{stu}_{\alpha \beta \alpha' \beta'}
+ \delta_{\tau_\alpha \tau_\beta} \delta_{\tau_{\alpha'} \tau_{\beta'}}
(1-\delta_{\tau_\alpha \tau_{\alpha'}})
c^s_{\alpha \beta \alpha' \beta'} } \nonumber \\
& & + \delta_{\tau_\alpha \tau_{\alpha'}} \delta_{\tau_\beta \tau_{\beta'}}
(1-\delta_{\tau_\alpha \tau_\beta})
c^{t1}_{\alpha \beta \alpha' \beta'}
+ \delta_{\tau_\alpha \tau_{\beta'}} \delta_{\tau_\beta \tau_{\alpha'}}
(1-\delta_{\tau_\alpha \tau_\beta})
c^{t2}_{\alpha \beta \alpha' \beta'}
\label{czerlegung}
\eea
with the isospin channel correlation functions
$c^{stu}$, $c^{s}$, $c^{t1}$, $c^{t2}$.

The one-body density matrices fulfill
\be
\rho_{\alpha \alpha'} = (\rho_{\alpha' \alpha})^* \; ,
\ee
while the isospin channel correlation functions fulfill
\be
c^{stu,s}_{\alpha \beta \alpha' \beta'}=
c^{stu,s}_{\beta \alpha \alpha' \beta'}=
c^{stu,s}_{\alpha \beta \beta' \alpha'}=
(c^{stu,s}_{\alpha' \beta' \alpha \beta})^*
\ee
and
\bea
c^{t1}_{\alpha \beta \alpha' \beta'}=
c^{t2}_{\beta \alpha \alpha' \beta'}=
c^{t2}_{\alpha \beta \beta' \alpha'}=
(c^{t1}_{\alpha' \beta' \alpha \beta})^* \; , \; \; \; \;
c^{t2}_{\alpha \beta \alpha' \beta'}=
(c^{t2}_{\alpha' \beta' \alpha \beta})^* \; .
\eea

We note in passing, that for our purpose the cartesian representation of
SU(2) is technically superior to the spherical one, because in the spherical
representation we have to consider 7 instead of 4 different
isospin channel correlation functions.

We can now explicitly carry out all summations over isospin indices
appearing in the NCBCD equations, which leads to a coupled system of
a single one-body equation of motion and four two-body equations of motion,
where all considered quantities now only depend on space-time degrees of
freedom.

With the abbreviations
\bea
U_{\alpha \alpha'}=\sum_{\lambda_1 \lambda_2} \rho_{\lambda_1 \lambda_2}
\left( 3{\langle}\alpha \lambda_2 |v_{t,u}|\alpha' \lambda_1{\rangle}
+ {\langle}\alpha \lambda_2 |v_{t,u}|\lambda_1 \alpha'{\rangle}
+{\langle}\alpha \lambda_2 |v_s|\alpha' \lambda_1{\rangle}_S \right) \; ,
\eea
\be
h_{\alpha \alpha'}=t_{\alpha \alpha'} + U_{\alpha \alpha'} \; ,
\ee
\bea
Q^=_{\alpha \beta \alpha' \beta'} &=&
\delta_{\alpha \alpha'} \delta_{\beta \beta'} +
\rho_{\alpha \alpha'} \delta_{\beta \beta'} +
\rho_{\beta \beta'} \delta_{\alpha \alpha'} \; , \\
Q^{\perp}_{\alpha \beta \alpha' \beta'} &=&
\rho_{\beta \alpha'} \delta_{\alpha \beta'} -
\rho_{\alpha \beta'} \delta_{\beta \alpha'}
\eea
we obtain the NCBCD equations for the isosymmetric, isodiagonal
pionic system:
\renewcommand{\baselinestretch}{1.0}
\small \normalsize
\bea
\lefteqn{ i\partial_t \rho_{\alpha \alpha'} =
\sum_\lambda \left( h_{\alpha \lambda} \rho_{\lambda \alpha'} -
h_{\lambda \alpha'} \rho_{\alpha \lambda} \right)  } \nonumber \\
& & + \sum_{\lambda_1 \lambda_2 \lambda_3} \left\{
{\langle}\alpha \lambda_1 |v_{t,u}| \lambda_2 \lambda_3{\rangle} \left(
c^{stu}_{\lambda_2 \lambda_3 \alpha' \lambda_1}
+ 2 c^{t1}_{\lambda_2 \lambda_3 \alpha' \lambda_1} \right)
+ {\langle}\alpha \lambda_1 |v_s| \lambda_2 \lambda_3{\rangle} \left(
c^{stu}_{\lambda_2 \lambda_3 \alpha' \lambda_1}
+ 2 c^s_{\lambda_2 \lambda_3 \alpha' \lambda_1} \right)
\right. \nonumber \\
& & \left. - {\langle}\lambda_1 \lambda_2 |v_{t,u}| \alpha'
\lambda_3{\rangle}
\left(
c^{stu}_{\alpha \lambda_3 \lambda_1 \lambda_2}
+ 2 c^{t1}_{\alpha \lambda_3 \lambda_1 \lambda_2} \right)
-{\langle}\lambda_1 \lambda_2 |v_s|\alpha' \lambda_3{\rangle} \left(
c^{stu}_{\alpha \lambda_3 \lambda_1 \lambda_2}
+ 2 c^s_{\alpha \lambda_3 \lambda_1 \lambda_2} \right) \right\}
\label{isoonebody}
\eea
and
\begin{eqnarray*}
\lefteqn{ i\partial_t \left[ \begin{array}{c}
c^{stu}_{\alpha \beta \alpha' \beta'} \\
c^s_{\alpha \beta \alpha' \beta'} \\
c^{t1}_{\alpha \beta \alpha' \beta'} \\
c^{t2}_{\alpha \beta \alpha' \beta'}
\end{array} \right] =
\sum_\lambda \left\{ h_{\alpha \lambda} \left[ \begin{array}{c}
c^{stu}_{\lambda \beta \alpha' \beta'} \\
c^s_{\lambda \beta \alpha' \beta'} \\
c^{t1}_{\lambda \beta \alpha' \beta'} \\
c^{t2}_{\lambda \beta \alpha' \beta'}
\end{array} \right]
+ h_{\beta \lambda} \left[ \begin{array}{c}
c^{stu}_{\alpha \lambda \alpha' \beta'} \\
c^s_{\alpha \lambda \alpha' \beta'} \\
c^{t1}_{\alpha \lambda \alpha' \beta'} \\
c^{t2}_{\alpha \lambda \alpha' \beta'}
\end{array} \right]
 - h_{\lambda \alpha'} \left[ \begin{array}{c}
c^{stu}_{\alpha \beta \lambda \beta'} \\
c^s_{\alpha \beta \lambda \beta'} \\
c^{t1}_{\alpha \beta \lambda \beta'} \\
c^{t2}_{\alpha \beta \lambda \beta'}
\end{array} \right]
- h_{\lambda \beta'} \left[ \begin{array}{c}
c^{stu}_{\alpha \beta \alpha' \lambda} \\
c^s_{\alpha \beta \alpha' \lambda} \\
c^{t1}_{\alpha \beta \alpha' \lambda} \\
c^{t2}_{\alpha \beta \alpha' \lambda}
\end{array} \right] \right\} }  \nonumber \\
& & + \sum_{\lambda_1 \lambda_2 \lambda_3 \lambda_4} \left\{
{\langle}\lambda_3 \lambda_4 |v_{t,u}| \lambda_1 \lambda_2 {\rangle}
Q^=_{\alpha \beta \lambda_3 \lambda_4}
\left[ \begin{array}{c}
(\rho_{20})_{\lambda_1 \lambda_2 \alpha' \beta'} \\
0 \\
\rho_{\lambda_1 \alpha'} \rho_{\lambda_2 \beta'} \\
\rho_{\lambda_1 \beta'} \rho_{\lambda_2 \alpha'}
\end{array} \right] \right. \nonumber \\
& & + {\langle}\lambda_3 \lambda_4 |v_s| \lambda_1 \lambda_2{\rangle}
Q^=_{\alpha \beta \lambda_3 \lambda_4}
\left[ \begin{array}{c}
(\rho_{20})_{\lambda_1 \lambda_2 \alpha' \beta'} \\
(\rho_{20})_{\lambda_1 \lambda_2 \alpha' \beta'} \\
0 \\
0
\end{array} \right] \nonumber \\
& & \left.
-{\langle}\lambda_3 \lambda_4 |v_{t,u}| \lambda_1 \lambda_2{\rangle}
Q^=_{\lambda_1 \lambda_2 \alpha' \beta'}
\left[ \begin{array}{c}
(\rho_{20})_{\alpha \beta \lambda_3 \lambda_4} \\
0 \\
\rho_{\alpha \lambda_3} \rho_{\beta \lambda_4} \\
\rho_{\alpha \lambda_4} \rho_{\beta \lambda_3}
\end{array} \right]
- {\langle}\lambda_3 \lambda_4 |v_s|\lambda_1 \lambda_2{\rangle}
Q^=_{\lambda_1 \lambda_2 \alpha' \beta'}
\left[ \begin{array}{c}
(\rho_{20})_{\alpha \beta \lambda_3 \lambda_4} \\
(\rho_{20})_{\alpha \beta \lambda_3 \lambda_4} \\
0 \\
0
\end{array} \right] \right\}  \nonumber \\
& & + \sum_{\lambda_1 \lambda_2 \lambda_3 \lambda_4} \left\{
{\langle}\lambda_3 \lambda_4 |v_{t,u}| \lambda_1 \lambda_2 {\rangle}
Q^=_{\alpha \beta \lambda_3 \lambda_4}
\left[ \begin{array}{c}
c^{stu}_{\lambda_1 \lambda_2 \alpha' \beta'} \\
c^{s}_{\lambda_1 \lambda_2 \alpha' \beta'} \\
c^{t1}_{\lambda_1 \lambda_2 \alpha' \beta'} \\
c^{t2}_{\lambda_1 \lambda_2 \alpha' \beta'}
\end{array} \right] \right. \nonumber \\
& & + {\langle}\lambda_3 \lambda_4 |v_s| \lambda_1 \lambda_2{\rangle}
Q^=_{\alpha \beta \lambda_3 \lambda_4}
\left[ \begin{array}{c}
c^{stu}_{\lambda_1 \lambda_2 \alpha' \beta'}
+ 2 c^s_{\lambda_1 \lambda_2 \alpha' \beta'} \\
c^{stu}_{\lambda_1 \lambda_2 \alpha' \beta'}
+ 2 c^s_{\lambda_1 \lambda_2 \alpha' \beta'} \\
0 \\
0
\end{array} \right]  \nonumber \\
& & \left.
-{\langle}\lambda_3 \lambda_4 |v_{t,u}| \lambda_1 \lambda_2{\rangle}
Q^=_{\lambda_1 \lambda_2 \alpha' \beta'}
\left[ \begin{array}{c}
c^{stu}_{\alpha \beta \lambda_3 \lambda_4} \\
c^{s}_{\alpha \beta \lambda_3 \lambda_4} \\
c^{t1}_{\alpha \beta \lambda_3 \lambda_4} \\
c^{t2}_{\alpha \beta \lambda_3 \lambda_4}
\end{array} \right]
- {\langle}\lambda_3 \lambda_4 |v_s|\lambda_1 \lambda_2{\rangle}
Q^=_{\lambda_1 \lambda_2 \alpha' \beta'}
\left[ \begin{array}{c}
c^{stu}_{\alpha \beta \lambda_3 \lambda_4}
+ 2 c^s_{\alpha \beta \lambda_3 \lambda_4} \\
c^{stu}_{\alpha \beta \lambda_3 \lambda_4}
+ 2 c^s_{\alpha \beta \lambda_3 \lambda_4} \\
0 \\
0
\end{array} \right] \right\}  \nonumber \\
& & + \sum_{\lambda_1 \lambda_2 \lambda_3 \lambda_4} \left\{
Q^{\perp}_{\alpha \lambda_1 \alpha' \lambda_3}
{\langle}\lambda_3 \lambda_4 |v_{t,u}|\lambda_1 \lambda_2{\rangle}
\left[ \begin{array}{c}
c^{stu}_{\lambda_2 \beta \lambda_4 \beta'}
+ 2 c^{t1}_{\lambda_2 \beta \lambda_4 \beta'} \\
0 \\
c^{stu}_{\lambda_2 \beta \lambda_4 \beta'}
+ 2 c^{t1}_{\lambda_2 \beta \lambda_4 \beta'} \\
0
\end{array} \right]  \right. \nonumber \\
& & + Q^{\perp}_{\alpha \lambda_1 \alpha' \lambda_3}
{\langle}\lambda_3 \lambda_4 |v_{t,u}|\lambda_2 \lambda_1{\rangle}
\left[ \begin{array}{c}
c^{stu}_{\lambda_2 \beta \lambda_4 \beta'} \\
c^{s}_{\lambda_2 \beta \lambda_4 \beta'} \\
c^{t1}_{\lambda_2 \beta \lambda_4 \beta'} \\
c^{t2}_{\lambda_2 \beta \lambda_4 \beta'}
\end{array} \right]
+ Q^{\perp}_{\alpha \lambda_1 \alpha' \lambda_3}
{\langle}\lambda_3 \lambda_4 |v_s| \lambda_1 \lambda_2{\rangle}_S
\left[ \begin{array}{c}
c^{stu}_{\lambda_2 \beta \lambda_4 \beta'} \\
c^{t2}_{\lambda_2 \beta \lambda_4 \beta'} \\
c^{t1}_{\lambda_2 \beta \lambda_4 \beta'} \\
c^{s}_{\lambda_2 \beta \lambda_4 \beta'}
\end{array} \right]  \nonumber \\
& & + Q^{\perp}_{\beta \lambda_1 \beta' \lambda_3}
{\langle}\lambda_3 \lambda_4 |v_{t,u}|\lambda_1 \lambda_2{\rangle}
\left[ \begin{array}{c}
c^{stu}_{\lambda_2 \alpha \lambda_4 \alpha'}
+ 2 c^{t1}_{\lambda_2 \alpha \lambda_4 \alpha'} \\
0 \\
c^{stu}_{\lambda_2 \alpha \lambda_4 \alpha'}
+ 2 c^{t1}_{\lambda_2 \alpha \lambda_4 \alpha'} \\
0
\end{array} \right]  \nonumber \\
& & + Q^{\perp}_{\beta \lambda_1 \beta' \lambda_3}
{\langle}\lambda_3 \lambda_4 |v_{t,u}|\lambda_2 \lambda_1{\rangle}
\left[ \begin{array}{c}
c^{stu}_{\lambda_2 \alpha \lambda_4 \alpha'} \\
c^{s}_{\lambda_2 \alpha \lambda_4 \alpha'} \\
c^{t1}_{\lambda_2 \alpha \lambda_4 \alpha'} \\
c^{t2}_{\lambda_2 \alpha \lambda_4 \alpha'}
\end{array} \right]
+ Q^{\perp}_{\beta \lambda_1 \beta' \lambda_3}
{\langle}\lambda_3 \lambda_4 |v_s| \lambda_1 \lambda_2{\rangle}_S
\left[ \begin{array}{c}
c^{stu}_{\lambda_2 \alpha \lambda_4 \alpha'} \\
c^{t2}_{\lambda_2 \alpha \lambda_4 \alpha'} \\
c^{t1}_{\lambda_2 \alpha \lambda_4 \alpha'} \\
c^{s}_{\lambda_2 \alpha \lambda_4 \alpha'}
\end{array} \right]  \nonumber \\
& & + Q^{\perp}_{\alpha \lambda_1 \beta' \lambda_3}
{\langle}\lambda_3 \lambda_4 |v_{t,u}|\lambda_1 \lambda_2{\rangle}
\left[ \begin{array}{c}
c^{stu}_{\lambda_2 \beta \lambda_4 \alpha'}
+ 2 c^{t1}_{\lambda_2 \beta \lambda_4 \alpha'} \\
0 \\
0 \\
c^{stu}_{\lambda_2 \beta \lambda_4 \alpha'}
+ 2 c^{t1}_{\lambda_2 \beta \lambda_4 \alpha'}
\end{array} \right]  \nonumber \\
& & + Q^{\perp}_{\alpha \lambda_1 \beta' \lambda_3}
{\langle}\lambda_3 \lambda_4 |v_{t,u}| \lambda_2 \lambda_1{\rangle}
\left[ \begin{array}{c}
c^{stu}_{\lambda_2 \beta \lambda_4 \alpha'} \\
c^{s}_{\lambda_2 \beta \lambda_4 \alpha'} \\
c^{t2}_{\lambda_2 \beta \lambda_4 \alpha'} \\
c^{t1}_{\lambda_2 \beta \lambda_4 \alpha'}
\end{array} \right]
+ Q^{\perp}_{\alpha \lambda_1 \beta' \lambda_3}
{\langle}\lambda_3 \lambda_4 |v_s|\lambda_1 \lambda_2{\rangle}_S
\left[ \begin{array}{c}
c^{stu}_{\lambda_2 \beta \lambda_4 \alpha'} \\
c^{t2}_{\lambda_2 \beta \lambda_4 \alpha'} \\
c^{s}_{\lambda_2 \beta \lambda_4 \alpha'} \\
c^{t1}_{\lambda_2 \beta \lambda_4 \alpha'}
\end{array} \right]  \nonumber \\
& & + Q^{\perp}_{\beta \lambda_1 \alpha' \lambda_3}
{\langle}\lambda_3 \lambda_4 |v_{t,u}|\lambda_1 \lambda_2{\rangle}
\left[ \begin{array}{c}
c^{stu}_{\lambda_2 \alpha \lambda_4 \beta'}
+ 2 c^{t1}_{\lambda_2 \alpha \lambda_4 \beta'} \\
0 \\
0 \\
c^{stu}_{\lambda_2 \alpha \lambda_4 \beta'}
+ 2 c^{t1}_{\lambda_2 \alpha \lambda_4 \beta'}
\end{array} \right]  \nonumber \\
& & \left. + Q^{\perp}_{\beta \lambda_1 \alpha' \lambda_3}
{\langle}\lambda_3 \lambda_4 |v_{t,u}| \lambda_2 \lambda_1{\rangle}
\left[ \begin{array}{c}
c^{stu}_{\lambda_2 \alpha \lambda_4 \beta'} \\
c^{s}_{\lambda_2 \alpha \lambda_4 \beta'} \\
c^{t2}_{\lambda_2 \alpha \lambda_4 \beta'} \\
c^{t1}_{\lambda_2 \alpha \lambda_4 \beta'}
\end{array} \right]
+ Q^{\perp}_{\beta \lambda_1 \alpha' \lambda_3}
{\langle}\lambda_3 \lambda_4 |v_s|\lambda_1 \lambda_2{\rangle}_S
\left[ \begin{array}{c}
c^{stu}_{\lambda_2 \alpha \lambda_4 \beta'} \\
c^{t2}_{\lambda_2 \alpha \lambda_4 \beta'} \\
c^{s}_{\lambda_2 \alpha \lambda_4 \beta} \\
c^{t1}_{\lambda_2 \alpha \lambda_4 \beta'}
\end{array} \right] \right\} \; .
\end{eqnarray*}
\bea
\label{isotwobody}
\eea
\renewcommand{\baselinestretch}{1.0}
\small \normalsize
\end{appendix}
\newpage
\newpage
\section{Figure Captions}
\newcounter{figno}
\begin{list}%
{\underline{fig.\arabic{figno}}:}%
{\usecounter{figno}\setlength{\rightmargin}{\leftmargin}}
\item
\label{onehart}
The Hartree-terms in the t-, u- (l.h.s.) and s-channel (r.h.s.);
dashed horizontal lines are spacelike, dashed vertical lines are timelike
$\sigma$-fields, solid lines represent pions.
\item
\label{onefock}
The Fock-terms in the t-,u- (l.h.s.) and s-channel (r.h.s.).
\item
\label{bildonecoll}
The term coupling to the two-body correlation
function; t-, u-channel (l.h.s.), s-channel (r.h.s.).
\item
\label{bildtwoborn}
Contributions to the Born-term.
\item
\label{bildtwogmat}
Resummation terms in vertical direction
with $Q^=$ in the incoming channel; t-,u-channel term (l.h.s.),
s-channel term (r.h.s.).
\item
\label{bildtworpat}
Resummation terms in horizontal direction
in the t-,u-channel; direct term (upper part), exchange term (lower part).
Note that $Q^\perp$ connects $\alpha$ and $\lambda_3$, $\alpha'$
and $\lambda_1$.
\item
\label{bildtworpas}
Resummation terms in horizontal direction
in the s-channel; direct and exchange term are identical.
\item
\label{bildene}
Single particle energy levels and occupation numbers
of the Hartree-Fock initialization for 60 (upper part) and 90 (lower part)
pions in a volume of $(10 \; {\rm fm})^3$ at various temperatures.
The symbols denote the position of the discrete energy levels.
\item
\label{bildavcorre}
Time-averaged correlation energy as a function
of the pion number at 100 MeV temperature for NCBCD, vertical and horizontal
approximation (upper part); magnification (lower part).
\item
\label{bildconv}
Change in total energy after switching on the
residual interaction as a function of $t_f$.
\item
\label{bildcorrconv}
Correlation energy with respect to $H$
as function of $t/t_f$ for different values of $t_f$.
\item
\label{bildabcd}
Change of the single contributions to the total
energy as function of $t_f$; {\bf A} kinetic energy,
{\bf B} mean-field-energy, {\bf C} total energy, {\bf D} correlation energy.
\item
\label{bild50150x}
$\int dy dz \rho_2(\vec{r}, \vec{r};\vec{r},\vec{r})$
(solid line),
$\int dy dz \rho_{20}(\vec{r}, \vec{r};\vec{r},\vec{r})$
(dashed line) and \\
$\int dy dz \rho(\vec{r};\vec{r}) \rho(\vec{r};\vec{r})$
(dotted line) as a function of $x$ at 50 MeV (upper part)
and 150 MeV (lower part) temperature for various densities.
\item
\label{bildvett}
$\int d^3 r
\rho_2 (\vec{r},\vec{r};\vec{r},\vec{r}) / \int d^3 r
\rho(\vec{r};\vec{r}) \rho(\vec{r};\vec{r})$
for various temperatures as a function of density.
\item
\label{bild3d50m10200m30}
Cut through the two-body density divided by the product of one-body
densities at $y_1=y_2=z_1=z_2=5 \; {\rm fm}$ at 50 MeV temperature and
30 pions in $(10 {\rm fm})^3$ (upper part) and at 200 MeV and 90 pions
(lower part); $\rho_{20}/\rho\rho$ (l.h.s.), $\rho_2/\rho\rho$ (r.h.s.).
\item
\label{bild50150m}
One-dimensional cut through the two-body density divided by the
product of
one-body densities at 50 MeV (upper part) and 150 MeV (lower part)
temperature and for various densities;
the relative distance is parameterized by $x$.
Solid line $\rho_2/\rho\rho$, dashed line $\rho_{20}/\rho\rho$.
\item
\label{bildfpeaks}
Relative correlation strength after switching on the
residual interaction with $t_f=160 {\rm fm}/c$.
\item
\label{bilddouble}
Correlation strength at 100 MeV temperature;
solid line with normal coupling, dashed line with doubled coupling.
\end{list}
\end{document}